\documentclass[twocolappendix]{emulateapj}

\usepackage{graphicx}
\usepackage{enumerate}
\usepackage{amsmath}
\usepackage{comment}
\usepackage{subfigure}
\usepackage{xcolor}

\begin{document}

\author{Juhan Kim\altaffilmark{1}, Yun-Young Choi\altaffilmark{2},
Sungsoo S. Kim\altaffilmark{3}, and
Jeong-Eun Lee\altaffilmark{3}}
\affil{
$^1$Center for Advanced Computation, Korea Institute for Advanced Study, Heogiro 85, Seoul 130-722, Korea\\
$^2$Department of Astronomy and Space Science, Kyung Hee University, Gyeonggi 446-701, Korea\\
$^3$School of Space Research, Kyung Hee University, Gyeonggi 446-701, Korea}

\title{Stochastic Model of the Spin Distribution of Dark Matter Halos}


\begin{abstract}
We employ a stochastic approach to probing the origin of the log-normal distributions of halo spin in N-body
simulations. After analyzing spin evolution in halo merging trees, it was found that a spin change can be
characterized by a stochastic random walk of angular momentum. Also, spin distributions generated by random
walks are fairly consistent with those directly obtained from N-body simulations. We derived a stochastic
differential equation from a widely used spin definition and measured the probability distributions of the derived
angular momentum change from a massive set of halo merging trees. The roles of major merging and accretion are
also statistically analyzed in evolving spin distributions. Several factors (local environment, halo mass, merging
mass ratio, and redshift) are found to influence the angular momentum change. The spin distributions generated in
the mean-field or void regions tend to shift slightly to a higher spin value compared with simulated spin
distributions, which seems to be caused by the correlated random walks. We verified the assumption of
randomness in the angular momentum change observed in the N-body simulation and detected several degrees of
correlation between walks, which may provide a clue for the discrepancies between the simulated and generated
spin distributions in the voids. However, the generated spin distributions in the group and cluster regions
successfully match the simulated spin distribution. We also demonstrated that the log-normality of the spin
distribution is a natural consequence of the stochastic differential equation of the halo spin, which is well described
by the Geometric Brownian Motion model.
\end{abstract}

\keywords{cosmology: theory --- halo spin --- large-scale structure of universe
--- methods: numerical --- methods: statistical --- random walk}

\maketitle

\section{Introduction}\label{sec:intro}
Primordial angular momentum is believed to decay with the expansion 
of the universe and the observed rotations of galaxies and clusters are believed 
to originate from tidal interactions with local environments (the tidal torque theory)
or the acquisition of orbital angular momentum from infalling matter (the merging theory). 
In the tidal torque theory \citep{hoyle49,peebles69,lee01,prociani02,lee07},
a collapsing cloud at the turn-around epoch may obtain a tidal torque
due to the misalignment between the local tidal shear and structural inertia tensor.
Consequently, the spin directions of halos may provide information on the nearby
mass distribution 
\citep{aragon-calvo14,zhang09, trowland13}, although some studies \citep{lee07,varela12}
have reported inconclusive or opposing results from observations.
On the other hand, the merging theory \citep{gardner01,vitvitska02,maller02}
focuses on the change in angular momentum in hierarchical clusterings
through the off-center infall of nearby objects/matter
to drive the sudden change in halo spin \citep{bett12}.
However, \cite{d'onghia07} argued that simulated
halos do not show any obvious correlation between 
spin and merger history.

Each theory views the same object from different angles.
A dark matter
halo is not static but dynamically evolves 
through the sometimes violent, sometimes quiet ingestion of nearby matter.
According to hierarchical clustering \citep{cole96,jiang14}, large 
halos grow in mass by assimilating nearby smaller halos. 
However, a small satellite halo tends to have a spiral infall motion 
with a finite 
orbital angular momentum around the central halo, which, after merging,
causes a net change in rotational angular momentum of the merging remnants.
This non-radial infall motion of satellites may be derived from the tidal torque
of anisotropic matter distributions around the central halo.
However, the acquisition of satellite orbital angular momentum 
described in the merging theory is analogous to
that of a mass shell in the tidal torque theory 
because they are based on the same 
ground: the tidal torque generated by local environments. 
The only difference is whether they consider the infalling source to be
a satellite or a mass shell.
By extending the original tidal torque theory of a spherical collapse
to the multiple-shell collapses at different epochs, the tension 
between the two theories
may be relaxed.

The distributions of halo spin have been reported to be well fitted by the 
log-normal function \citep{gardner01,bailin05,bullock01,munoz-cuartas11},
with small deviations depending on the definition of the halo
samples \citep{bett07,knebe08,antonuccio-delogu10}.
Researchers have found that massive cluster halos are not yet relaxed from 
recent mergers and are responsible for the 
over-population of the high-$\lambda$ tail compared to the log-normal distribution
\citep{bett07,hahn07}.

\cite{antonuccio-delogu10} explored the origin of the log-normality of spin distributions
based on the stochastic variables of halo angular momentum ($\boldsymbol{J}$), total energy ($E$), 
and mass ($M$).
Theirs was the first attempt of its kind to address this issue. They introduced stochasticity
between the two terms ${|\boldsymbol{J}|}/M^{5/2}$ and $E$ and
argued that deviations from the log-normal functional form
are mainly due to correlations between the two terms.
However, they did not provide a physical or probable reason for the
stochasticity between the two terms.

In the present study, we analyze the origin of the halo spin distribution.
Using simulated mass histories of individual halos and the stochastic 
differential equation of the angular momentum change,
we model the spin evolution using the {\it stochastic model},
distinguishing between the roles of major merging and accretion in shaping 
the spin distributions in various situations 
\citep{hetznecker06,antonuccio-delogu02,gardner01,lemson99}.

This paper is organized as follows. 
In section~\ref{sec:spindef}, we describe the stochastic for
spin evolution. Then, in section~\ref{sec:data}, 
we describe the simulation data, merging trees, and local environments of 
a halo. 
From the simulated merging history, 
we measure distributions of angular momentum change with respect to halo mass, 
spin, local environments, infall 
mass ratio, and redshift in section~\ref{sec:change}. 
In section~\ref{sec:spin}, 
using the simulated mass histories of individual halos, 
we randomly generate spin changes 
and compare the spin distributions of the stochastic model 
and $N$-body simulations.
In section~\ref{sec:markov}, we verify how a random-walk process
produces the log-normal distribution of the halo spin.
We summarize the results of our study in section~\ref{sec:concl}

All of the logarithms used in the equations and figures in this study are base 10.

\section{Analytic Descriptions}\label{sec:spindef}
\subsection{Definition of Spins}
\cite{peebles69} first proposed a dimensionless parameter to quantify 
the rotation of a cosmic object or halo as
\begin{equation}
\lambda = {{\sqrt{E} |\boldsymbol{J}| } \over {GM^{5/2}} },
\end{equation}
where $\lambda$ is the spin, $E$ is the total energy, $M$ is the virial mass, 
and $\boldsymbol{J}$ is the rotational angular momentum of a halo. 
However, measuring halo potential is practically bothersome, and
the potential of an open system is sometimes ill-defined.
Therefore,
another definition was proposed by \cite{bullock01}, which reads as
\begin{equation}
\lambda^\prime_o = {{|\bf{J}| }\over \sqrt{2} MRV},
\label{spinp}
\end{equation}
where
$R$ is the virial radius, and $V$ is the virial circular velocity.
The virial radius can be derived from the definition of virial mass as
\begin{equation}
M = {4\pi\over3} R^3 \rho_c(z)\Delta_{\rm vir}(z),
\label{Mvir}
\end{equation}
where $\rho_c$ is the critical density at redshift $z$,
 $\Delta_{\rm vir} \simeq (18\pi^2 + 82x-39x^2)$,
and $x\equiv \Omega(z)-1$ for a flat universe \citep{bryan98}.
These widely used spin parameters are equivalent to $\lambda$ in 
the case of an idealized halo;
halos are fully virialized and their density profiles follow 
the isothermal model.
However, simulated halos may show a significant discrepancy
between $\lambda$ and $\lambda_0^{\prime}$ because most are not fully 
virialzed and their radial profiles do not follow the isothermal model. 

The gravitational potential of a halo should be measured using
the same finite softening length ($\epsilon$) as adopted in $N$-body simulations
because the chosen potential estimation strongly depends on the potential model:
the Newtonian ($1/r$) or Plummer ($1/\sqrt{\epsilon^2+r^2}$) model.
The difference increases at higher redshift \citep{ahn14}.
To comply with the simulated potential, we measure the halo potential using
\begin{equation}
\Phi(r) = - { Gm_im_j\over \sqrt{ \epsilon^2+r^2}},
\end{equation}
where $\epsilon = 36 ~ h^{-1} {\text{kpc}}$.
Therefore, great care must to be taken when interpreting the halo spin value.

In this study, we adopt a new spin parameter $\lambda^{\prime}$
\citep{ahn14}, 
\begin{eqnarray}
\lambda^\prime &=& f(z) \lambda_0^\prime\\
&=&  \left({0.26 \over \sqrt{1+z} }\right)^{1/6} \lambda_0^\prime,
\end{eqnarray}
which is introduced to correct for all possible systematics (see \citealt{ahn14}).
We call $f(z)$ a correcting factor.  
Hereafter, whenever we refer to a halo spin,
we mean the corrected spin and use $\lambda$ to denote it.
Of course, one may return to $\lambda_0^\prime$ simply by dividing the 
spin value by $f(z)$.

\subsection{Stochastic Model: Random Walk of Halo Spin}
From equation~(\ref{spinp}), 
we derive a stochastic differential equation of spin change as
\begin{equation}
{d\log_{10}\lambda\over d\log_{10} M} = {d\log_{10} |{\boldsymbol{J}}|\over d\log_{10} M}  + \alpha (z)- {
5\over3},
\label{excur}
\end{equation}
where 
\begin{equation}
\alpha(z) \equiv  {1\over6}  {d\log_{10} (\rho_c(z) \Delta_{\rm vir})\over d\log_{10} M}
+ {d\log_{10} f(z)\over d\log_{10} M}.
\label{cosmos}
\end{equation}
Here, $\alpha(z)$ is a cosmological effect depending on the redshift and model parameters. 
However, its contribution is relatively trivial compared with the 
constant factor of $-5/3$ in equation~(\ref{excur})
unless $|d\log_{10} M|\lesssim 10^{-2}$.
This implies that the background history of the universe does not much affect
the evolution of the spin parameter (Lemson \& Kauffmann 1999; Cervantes-Sodi et al. 2008). 

This differential equation was derived to model the spin change as a function of
the mass event (mass-merging or mass-loss event)
accompanying a change in angular momentum amplitude $|\boldsymbol{J}|$.
Here, we introduce a Markov chain process
of spin evolution as
\begin{equation}
{\Delta \log_{10} \lambda \over \Delta\log_{10} M }
= D(M,\lambda,z,{\rm etc.}) -D_c,
\label{eq:walk}
\end{equation}
where $D_c \equiv 5/3-\alpha(z)$ and the angular momentum change rate is defined as
\begin{equation}
D(M,\lambda,z,{\rm etc.}) \equiv  {\Delta\log_{10}|{\boldsymbol{J}}|\over \Delta\log_{10} M}. \label{eq:D} 
\end{equation} 
Since $\alpha(z)$ is relatively small, we will not explicitly
mention it, even though we include this
contribution in the subsequent calculations. 
We stress the importance of $\alpha(z)$ later in Appendix~\ref{sec:alpha}.

Throughout this stochastic approach, 
our primary assumption is that a spin change, $\Delta \log_{10}\lambda$, 
does not depend on the past history of a halo but can be completely
described by the current physical status of the halo;
the spin change is determined by the stochastic probability
of angular momentum change, $P(D)$,
which will later be shown to be a function of $M$, $\lambda$, 
$\Delta \log_{10}M$, $z$,
and local environments. 
Later, we will prove that if $P(D)$ is a Gaussian distribution,
then the resulting spin distribution follows the log-normal shape.
In the stochastic equation, there is one independent variable, 
$\Delta\log_{10}M$, which should be determined outside of the model. 
The halo mass history could be provided by $N$-body simulations or 
by an extended Press \& Schechter (EPS) formalism. However,
in this analysis, we limit our study to the simulated mass 
histories.
Hereafter, we define a major merger event and an accretion (or minor-merger) 
event as mass changes of $\Delta \log_{10} M \ge 0.1$
and $0 < \Delta \log_{10} M <0.1$, respectively.

\section{Data}\label{sec:data}
\subsection{Simulations}
We run a series of cosmological $N$-body simulations with
$2048^3$ particles using the GOTPM code \citep{dubinski05}.
Four of the simulations are basically identical but were run with different sets of initial conditions.
The simulations have a cubic simulation volume with
a side length of $L_{\rm box}=737.28~h^{-1}{\rm Mpc}$.
The cosmological model adopted in the simulations
is a WMAP 5-year cosmology. The fractions of matter, baryonic matter, and dark energy
are 0.26, 0.044, and 0.74, respectively. The initial power spectrum is obtained from
the CAMB Source package \footnote{http://camb.info/sources}.
Table \ref{sim} summarizes several characteristics of the simulations used in this paper.

\begin{deluxetable} {c c c c c c c c c } 
\tablecaption{Simulation characteristics}
\tablehead{
\colhead{$N_p$}
&\colhead{$L_{\rm box}$}
&\colhead{$z_i$\tablenotemark{a}} 
&\colhead{$N_{\rm step}$} 
&\colhead{$\Omega_m$} 
&\colhead{$\Omega_b$}
&\colhead{$\Omega_\Lambda$} 
&\colhead{$h$\tablenotemark{b}} 
&\colhead{$b_8$\tablenotemark{c}}
}
\startdata
$2048^3$ &737.24 & 120 & 3000 & 0.26 & 0.044 & 0.74 & 0.72 & 1.26
\enddata
\tablenotetext{a}{Starting redshift of the simulation}
\tablenotetext{b}{Hubble expansion parameter divided by 100 
kms$^{-1}$Mpc$^{-1}$}
\tablenotetext{c}{Bias factor at $R=8~h^{-1}\rm Mpc$}
\label{sim}
\end{deluxetable}

We applied the Friend-of-Friend (FoF) method to identify groups of particles
with a percolation linking length, $l_{\rm FoF}$, which is usually (also in this paper) 
set to be 0.2 times
the mean particle separation. This ensures that the identified halos to have a
mean over-density of 180, satisfying the cosmic virialization conditions 
in the spherical top-hat collapse model.
The minimum mass of a halo identified with 30 member particles is about $10^{11}~h^{-1} {\rm M_\odot}$.
Data on the member particles of each halo are stored at 44 redshifts
during the simulation run.
Table \ref{mertable} contains some basic information on
the merging data extracted from these $N$-body simulations,
including the simulation step number ({\it first column}),
redshift ({\it second}), time between merging steps ({\it third}),
number of halos ({\it fourth}), number of seed halos ({\it fifth}),
 and mean halo number density ({\it last}).
Here, seed halos are introduced to define a local density which
will be described in detail in section~\ref{localden}.

\begin{deluxetable}{c|llrrc}
\tablecaption{Basic information on the merging data}
\tablehead{
\colhead{step\tablenotemark{a}} 
&\colhead{$z$} 
&\colhead{$\Delta t$\tablenotemark{b}} 
&\colhead{$N_{\rm samp}$\tablenotemark{c}}
&\colhead{$N_{\rm bkg}$\tablenotemark{d}}
&\colhead{$\bar{n}$\tablenotemark{e}} 
}
\startdata
250 &       10  &   -- & 174 & 174 & $4.3\times 10^{-7}$ \\
278 &        9  & $125.7895$  & 7,317 & 713  & $4.6\times 10^{-6}$ \\
353 &        7  & $336.9363$  & 264,329 & 47,823 & 0.00016 \\
407 &        6  & $242.5939$  & 1,135,317 & 3,735 & 0.00071 \\
440 &      5.5  & $148.2521$  & 2,142,976 & 11,608 &  0.0013 \\
479 &        5  & $175.2068$  & 3,841,831 & 34,147 &  0.0024 \\
501 &      4.8  & $ 98.8345$  & 5,025,833 & 56,128 &  0.0031 \\
525 &      4.5  & $107.8197$  & 6,471,073 & 89,473 &   0.004 \\
551 &      4.3  & $116.8045$  & 8,190,543 & 139,552 &  0.0051 \\
580 &        4  & $130.2820$  & 10,239,195 & 213,106 &  0.0064 \\
605 &      3.8  & $112.3119$  & 12,072,304 & 292,010 &  0.0075 \\
633 &      3.6  & $125.7898$  & 14,159,663 & 397,941 &  0.0088 \\
662 &      3.4  & $130.2818$  & 16,313,976 & 526,000 &    0.01 \\
695 &      3.2  & $148.2521$  & 18,714,095 & 692,528 &   0.012 \\
731 &        3  & $161.7294$  & 21,221,990 & 893,728 &   0.013 \\
771 &      2.8  & $179.6993$  & 23,841,886 & 1,137,628 &   0.015 \\
815 &      2.6  & $197.6691$  & 26,477,254 & 1,419,370 &   0.017 \\
865 &      2.4  & $224.6239$  & 29,135,986 & 1,746,844 &   0.018 \\
920 &      2.2  & $247.0869$  & 31,672,320 & 2,105,555 &    0.02 \\
983 &        2  & $283.0264$  & 34,107,018 & 2,500,701 &   0.021 \\
1018 &      1.9  & $157.2365$  & 35,262,139 & 2,708,351 &   0.022 \\
1055 &      1.8  & $166.2222$  & 36,346,533 & 2,918,928 &   0.023 \\
1095 &      1.7  & $179.6993$  & 37,374,958 & 3,132,572 &   0.023 \\
1138 &      1.6  & $193.1770$  & 38,330,932 & 3,347,949 &   0.024 \\
1185 &      1.5  & $211.1466$  & 39,224,698 & 3,565,667 &   0.024 \\
1235 &      1.4  & $224.6238$  & 40,019,836 & 3,777,393 &   0.025 \\
1290 &      1.3  & $247.0869$  & 40,747,200 & 3,985,683 &   0.025 \\
1350 &      1.2  & $269.5488$  & 41,384,587 & 4,188,379 &   0.026 \\
1415 &      1.1  & $292.0116$  & 41,921,471 & 4,379,074 &   0.026 \\
1487 &        1  & $323.4587$  & 42,370,158 & 4,561,970 &   0.026 \\
1567 &      0.9  & $359.3986$  & 42,728,496 & 4,731,685 &   0.027 \\
1656 &      0.8  & $399.8306$  & 42,996,082 & 4,887,605 &   0.027 \\
1754 &      0.7  & $440.2634$  & 43,161,903 & 5,027,849 &   0.027 \\
1866 &      0.6  & $503.1586$  & 43,252,426 & 5,155,748 &   0.027 \\
1992 &      0.5  & $566.0531$  & 43,252,029 & 5,264,860 &   0.027 \\
2136 &      0.4  & $646.9167$  & 43,180,668 & 5,362,503 &   0.027 \\
2302 &      0.3  & $745.7516$  & 43,035,938 & 5,443,524 &   0.027 \\
2496 &      0.2  & $871.5429$  & 42,835,401 & 5,513,029 &   0.027 \\
2725 &      0.1  & $1028.778$  & 42,586,126 & 5,570,734 &   0.027 \\
2776 &     0.08  & $229.1170$  & 42,533,514 & 5,581,988 &   0.027 \\
2856 &     0.05  & $359.3983$  & 42,446,174 & 5,597,921 &   0.026 \\
2941 &     0.02  & $381.8615$  & 42,359,628 & 5,611,664 &   0.026 \\
3000 &     0     & 268.061 & 42,298,156 & 5,621,397 &   0.026 
\enddata
\tablenotetext{a}{Simulation time step in units of Myr}
\tablenotetext{b}{Time interval between steps}
\tablenotetext{c}{Total number of halos with members equal to or larger than 30 simulation particles}
\tablenotetext{d}{Number of background halos needed to trace density field}
\tablenotetext{e}{Mean number density of halos in units of $h^3$Mpc$^{-3}$}
\label{mertable}
\end{deluxetable}

\subsection{Main Merging Tree}
To build halo merging trees, we used information for member particles 
of each FoF halo.
If a ``majority'' of member particles of a halo are also members of
a halo at the previous time step, then we link them 
with a mainr-merging tree line.
We call the former halo an ``ancestor'' and the latter a ``major descendent.''
If a halo has no major descendent, then we terminate the line.
The chains of these merging lines consist of a merging tree 
with which we are able to trace the
histories of halo angular momentum, mass, and other physical quantities.
One should note that
the main merger tree is a trimmed version of the merging tree widely quoted
in other studies.

Note that, on average, about 25\% of halos experience 
a mass loss in a time step of this study, 
which means that its major descendent (or successor) becomes less massive.
This happens when a halo undergoes a violent merging process 
and is broken into multiple less-massive remnants or when a halo 
passes another and is stripped of some member particles.

\subsection{Spin Evolution of Simulated Halos}
\label{localden}
\begin{figure}[tp]
\epsscale{1.}
\plotone{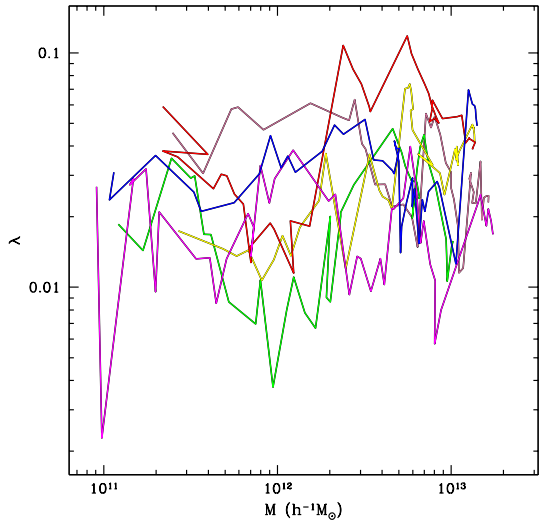}
\caption{
Several examples of spin evolution of simulated halos.
Each color represents the spin trajectory of a single main-merging tree. 
For the $x$ axis, we use halo mass rather than time or redshift.
Halos at $z=0$ are chosen with mass  
$10^{13}~h^{-1}{\rm M_\odot} \le M < 2\times 0^{13}~h^{-1}{\rm M_\odot}$.
}
\label{fig1}
\end{figure}

Now, let us look at the spin evolution of simulated halos
and their main-merging trees.
In Figure~\ref{fig1}, we show several typical examples
selected from the simulations. There are several significant spin decreases to
less than $\lambda= 0.01$ or increases to greater than $0.07$,
but halo spins immediately bounce back
to settle over the range of $0.01\lesssim \lambda \lesssim 0.07$.
Here, we suppose that spin change is random, but
there is a hidden mechanism that regulates the spin value to lie in a certain 
range.

A sudden massive merger event may cause a halo to rotate quickly, resulting in
a substantial loss of angular momentum due to the ejection of high-angular 
momentum material through violent relaxation.
A smaller halo may pass a nearby larger halo 
(being attached when they encounter but separated afterward),
after which point the larger halo experiences a sudden increase and decrease
in spin value.
Of course, during the encounter, there could be an exchange of finite angular 
momentum and the host may not fully recover its original angular momentum.
As stated before, several events of mass loss events due to the 
separation of satellites
can be seen in Figure~\ref{fig1} (a line segment moving to the left). 

\subsection{Local Environment}
We define a local density measured from 10 nearest neighbor halos 
as a halo environment.
The halos tracing the local density, called ``background halos'', are all halos 
with mass greater than 
$M_s = 10^{12} ~h^{-1}{\rm M_\odot}$ (equal to the sum of 300 particle masses) at $z<7$.
However, at higher redshifts ($z>7$), the number of massive halos is much 
lower and, therefore, we reduce the mass limit
to $M_s = 2\times 10^{11} h^{-1}{\rm M_\odot}$ at $7\leq z<9$ and even 
further to $M_s =  10^{11} h^{-1}{\rm M_\odot}$ at $z\ge 9$.

The variable-width spline kernel is adopted to adaptively resolve
high-density regions as
\begin{equation}
\rho^i = \sum_{j=1}^{N_{\rm n}=10} W(|{{\boldsymbol{ r}}^{ij}}|/ h^i),
\end{equation}
where ${\boldsymbol{ r}}^{ij}$ is the distance from the $i$th halo 
to the $j$th background halo, and $h^i$ is the distance to the tenth nearest neighbor.
The density measure was first used with 20 neighbors in Park et al. (2007).
The kernel size varying with environments helps dense regions 
not to be oversmoothed. 

The spline kernel has the form 
\begin{equation*}
W(x) = {8\over\pi}
\begin{cases}
1-6x^2 + 6x^3 & \text{if } 0\le x \le {1\over2},\\
2(1-x)^3 & \text{if } {1\over2} < x \le 1, \text{ and}\\
0 & \text{if }  x>1,
\end{cases}
\end{equation*}
where $x\equiv r/h$.
The median radius of the kernel is $h=7.5$ h$^-1$Mpc.

The local environment is parameterized by
\begin{equation}
\Delta\rho_{10}(r) \equiv {\rho_{10}(r)\over \bar\rho}
\end{equation}
where $\bar\rho$ is the mean density of the background halos, which
is measured to be 
$3.54\times 10^{-3}(\rm h^{-1}\rm Mpc)^{-3}=(6.56 \rm h^{-1}\rm Mpc)^{-3}$.
If a region has a density of $\Delta\rho_{10}<0.7$, 
then we refer to it as an under-dense or 
void region.
Regions $0.7\le\Delta\rho_{10}<2$, $2\le\Delta\rho_{10}<10$, $10\le\Delta\rho_{10}<100$,
and $\Delta\rho_{10}>100$ are referred to as mean-field, group, cluster,
and highly clustered regions, respectively.

\section{Probability Distribution of Angular Momentum Change, $P(D)$}
\label{sec:change}

\subsection{General Descriptions of $P(D)$}
\begin{deluxetable}{c | c c c c c}
\tablecaption{Parameter ranges\tablenotemark{a} applied to the measurement of $P(D)$}
\tablehead{
\colhead{id} 
&\colhead{halo mass ($M_{12}$)} 
&\colhead{$\lambda$} 
&\colhead{$\Delta\log_{10}M$}
&\colhead{$\Delta\rho_{10}$}
&\colhead{$z$}
}
\startdata
1 & 0.1 & 0.004 & -0.3 & 0.7 & 5\\
2 & 0.3 & 0.006 & -0.1 & 2 & 4\\
3& 0.5 & 0.0085 & -0.07 & 10 & 3\\
4 & 0.8 &  0.01 & -0.05 & 100 & 2\\
5 & 1 &  0.015 & -0.03 & $\infty$ & 1.5\\
6 & 2 & 0.02 & -0.02 & - & 1\\
7 & 4 & 0.025 & -0.01 & - & 0.6\\
8 & 5 & 0.03 & -0.0065 & - & 0.4\\
9 & 7 & 0.035 & -0.004 & - & 0.2\\
10 & 10 & 0.038 & 0 & - & 0.0 \\
11 & 20 & 0.04 & 0.004 & - & -\\
12 & 50 & 0.043 & 0.0065 & - & - \\
13 & 500 & 0.045 & 0.008 & - & - \\
14 &5000 & 0.047 & 0.01 & - & - \\
15 & $\infty$  & 0.05 & 0.015 & - & - \\
16 & - & 0.053 & 0.02 & - & -\\
17 & - & 0.056 & 0.03 & - & -\\
18 & - & 0.06 & 0.05 & - & - \\
19 & - & 0.063 & 0.08 & - & - \\
20 & - & 0.067 & 0.1 & - & -\\
21 & - & 0.07 & 0.15 & - & -\\
22 & - & 0.072 & 0.2 & - & - \\
23 & - & 0.075 & 0.3 & - & - \\
24 & - & 0.078 & 0.7 & - & - \\
25 & - & 0.08 & $\infty$ & - & -\\
26 & - & 0.09 & - & - & -\\
27 & - & 0.1 & - & - & -\\
28 & - & 0.11 & - & -& -\\
29 & - & 0.12 & -&-&-\\
30 & - & 0.13 & - &-&-\\
31 &-& 0.15 & -&-&-\\
32 &-&0.17 &-&-&-\\
33 &-&0.2&-&-&-\\
34&-&0.25&-&-&-\\
35&-& 0.3 & -&-&-\\
36&-& 0.4 & -&-&-\\
37 & - & 0.5 &-&-&-\\
38 &-& 0.7 &-&-&-\\
39 & - & 1.2 &-&-&-\\
40 & -& $\infty$ &-&-&-
\enddata
\tablenotetext{a}{The table lists the upper bound of the parameter range
except the redshift where we instead write the lower bound. Then, a parameter range
is $p_{i-1} \le {P} < p_i$, where $P$ is a parameter, and
$p_i$ is a value written in the $i$th row.
For $p_0$, the lowest possible value is assumed.}
\label{paratable}
\end{deluxetable}

We measured the distribution of angular momentum changes, $P(D)$,
as a function of $M$, $\lambda$, $\Delta\log_{10} M$, $z$, 
and $\Delta\rho_{10}$.
First, we divided all of the merging events into five-dimensional subsamples 
stretching over the entire parameter space, and
measured the distribution of $D$ in each subsample.
The $D$ was derived
by estimating the changes in mass and the angular momentum (Eq.~\ref{eq:D}).
Between merging steps ($t_1$ and $t_2$), 
the log mass change would be
$\log_{10}M_2 - \log_{10}M_1$ and the angular momentum change
in the log scale is obtained as
$\log_{10} |\boldsymbol{J}_2| - \log_{10}|\boldsymbol{J}_1|$,
where subscripts 1 and 2 indicate the time step.
Then, we fit the 
distribution with a bimodal Gaussian function as
\begin{eqnarray}
\nonumber
P&(&D;\mu_1,\sigma_1, \mu_2,\sigma_2) = 
{f_1\over \sqrt{2\pi\sigma_1^2}} \exp\left[{-{1\over2}{{(D-\mu_1)}^2\over\sigma_1^2}}\right] \\
&&+ {1-f_1\over \sqrt{2\pi\sigma_2^2}} \exp\left[{-{1\over2}{{(D-\mu_2)}^2\over\sigma_2^2}}\right] 
\end{eqnarray}
where $\mu$ is the Gaussian mean, and $\sigma$ is the standard deviation.
The overall mean ($\mu_D$) and standard deviation ($\sigma_D$) are obtained as 
\begin{eqnarray}
\mu_D &=& f_1 \mu_1 +(1-f_1) \mu_2 \\
\sigma_D &=& f_1 \left(\sigma_1^2 + d_1^2\right) 
 + (1-f_1)\left( \sigma_2^2+d_2^2\right),
\end{eqnarray}
where $d_1\equiv \mu_1 - \mu_D$ and $d_2 \equiv \mu_2-\mu_D$.

\begin{figure}[tp]
\plotone{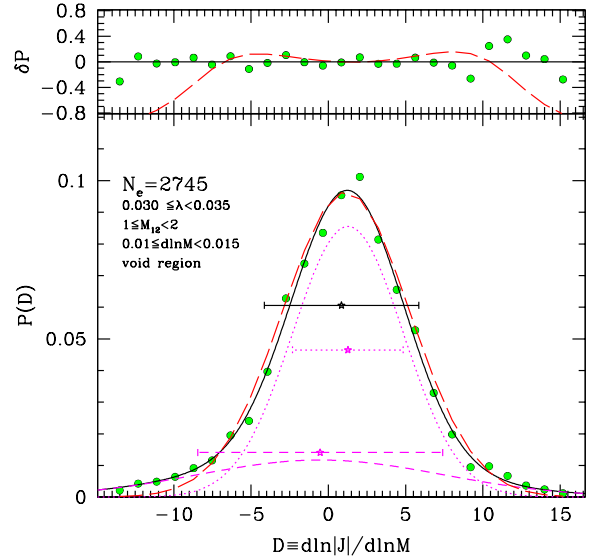}
\caption{Typical example of a distribution of $D$. In this plot, we select a 
sample with 
$10^{12} \leq M/(h^{-1}{\rm M_\odot}) <2\times 10^{12}$, $0.03\leq\lambda<0.035$, $0.01 \leq \Delta\log_{10} M < 0.015$, 
and $\Delta\rho_{10} <0.7$ 
in the redshift interval of $0 \leq z <0.2$.
A total of 2745 mass events are recorded in these parameter spaces.
Bottom:
The solid line with symbol and error bar shows the least $\chi^2$ fit, mean, and 
standard deviation of the distribution. Also, we plot the mean and 
standard deviation of each Gaussian component using dotted (major component) 
and dashed (minor component) lines. 
Their corresponding Gaussian components are shown using the same line styles.
The long dashed line is a single Gaussian fit to the symbols.
Top:
the deviations of $P(D)$ with respect to the bimodal Gaussian fit are shown
with circles. 
}
\label{fig2}
\end{figure}

Figure \ref{fig2} shows an example of the probability distribution of angular 
momentum change in a void sample ($\Delta \rho_{10} < 0.7$) with
parameter ranges $0.03\le\lambda<0.035$, 
$0.01\le \Delta\log_{10}M<0.015$, and $1\le M_{12}<2$,
where $M_{12}\equiv M/(10^{12}~ h^{-1}{\rm M_\odot})$.
A fitting to the bimodal Gaussian is shown by a solid line. 
The star symbol with a solid error bar
in  the middle of the plot represents the mean with 1--$\sigma$ standard 
deviation, while the lower two error bars (dotted and dashed) mark those of each Gaussian component.
Also, we overplot the single Gaussian fit (long dashed curve),
which shows a poorer fit to the fat tails on both sides (more easily seen in the upper panel).
Spin distributions generated based on the single Gaussian fitting function
show large deviations from simulated distributions, while the bimodal Gaussian has a better result
in describing simulated spin distribution at most redshifts.

There are several points to note in this plot. 
First, the distribution around a peak is well described by a single 
major Gaussian,
while the fat tails on both sides require an additional minor broad Gaussian.
Second, the peak positions of the Gaussian components are not well
separated from each other, implying that the combined distribution is somewhat 
symmetric. 
Most subsamples show this kind of distribution, although several show 
significant deviations from a symmetrical shape.
Such asymmetric distributions of $D$ may play a crucial role in distorted 
cluster spin distributions from 
a log-normal form, which will be discussed in subsequent sections.

\subsection{Comparison of $P(D)$ among Halo Subsamples}
\begin{figure}[tp]
\plotone{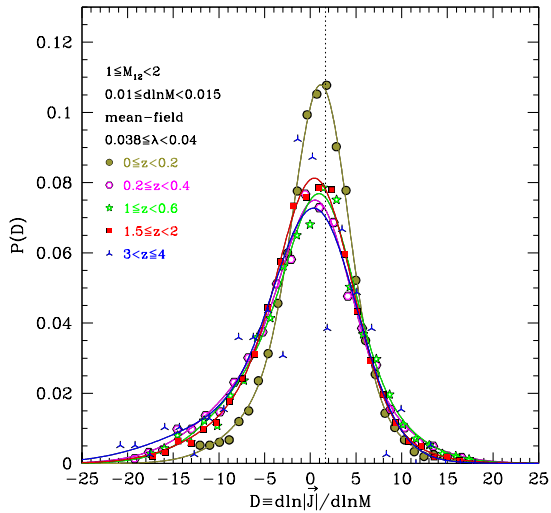}
\caption{
Redshift evolution of $P(D)$ for a spin range of 
$0.038 \le \lambda <0.04$ (in a moderately rotating case). 
Other parameter ranges of the sample are shown in the legend.
The vertical dotted line marks $D=D_c$ as a reference.
}
\label{fig3}
\end{figure}

In Figure \ref{fig3}, we show the redshift dependence of $P(D)$ in halo subsamples for a small mass
infall ratio ($0.01\le \Delta\log_{10} M < 0.015$) and an average spin 
value ($0.038\le \lambda < 0.04$) 
in the mean field ($0.7\le \Delta\rho_{10} < 2$).
As the redshift approaches zero,
the standard deviation of the distribution becomes narrower and the mean 
value becomes larger.
From this behavior,
we understand that the infall direction 
is more likely to be random at higher redshift because $\mu_D \simeq 0$
(see the distribution color-coded in blue at redshift between $z=$ 3 and 4).
At a lower redshift, angular momentum changes tend to be larger, suggesting that
the orbital angular momentum of infalling matter tends
to be a bit more aligned with (or positively biased to)
the halo rotational axis (also see Appendix \ref{appen1}).
If $D\simeq D_c$ ({\it vertical dotted line}), then we expect that
there would be no substantial change in the spin value.

\begin{figure}[tp]
\plotone{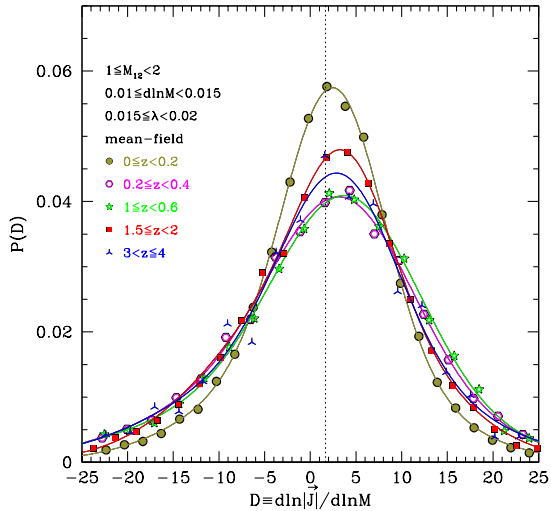}
\caption{
Same as Fig. \ref{fig3} but for low-spin sample of $0.015\le \lambda <0.02$,
which is used to isolate the spin effect on the angular momentum change.
}
\label{fig4}
\end{figure}

Figure \ref{fig4} shows a similar plot, but this time
with slowly rotating halo samples in the accretion mode.
All distributions have $\mu_D > D_{\rm c}$ indicating that
low-spin (or slow-rotating) halos are highly likely to increase in spin value.
The distribution spread is larger than in the mild-rotating samples 
(see Fig. \ref{fig3}),
implying that the spin change is larger for slow-rotating halos.
This is because incoming orbital angular momentum may
overwhelm the previous small rotational angular momentum. 

\begin{figure}[tp]
\plotone{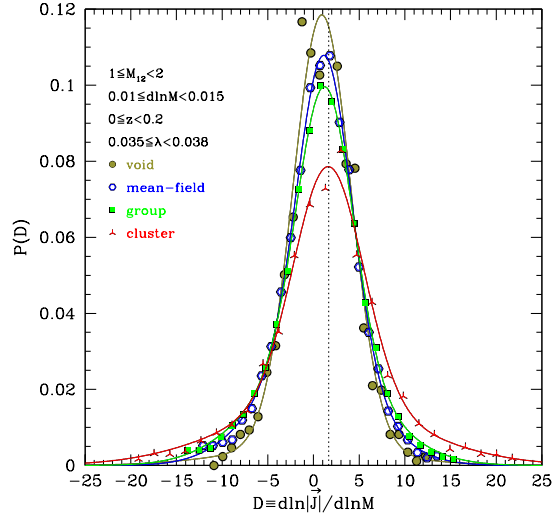}
\caption{
Environmental dependence of $P(D)$ for the moderately rotating halo samples
used in Fig.~\ref{fig3}.
}
\label{fig5}
\end{figure}

Next, we investigate the effect of environment on $D$.
Figure \ref{fig5} shows that
halos in cluster environments experience stronger 
angular momentum changes inferred from larger tails 
(or higher probability of large-$D$ changes).
This means that the angular momentum changes are statistically greater than 
those in lower-density regions, or that biased (or systematic) accretion or high-velocity mass infall is more frequent in cluster regions.
Also, one should note that each Gaussian component
has a similar peak position in most of the fittings.
However, when we select a sample of 
higher local density, the centers of two Gaussian components
begin to separate from each other. 
This effect on spin distribution will be described in detail in 
section \ref{sec:asym} and \ref{sec:env}.

\begin{figure}[tp]
\plotone{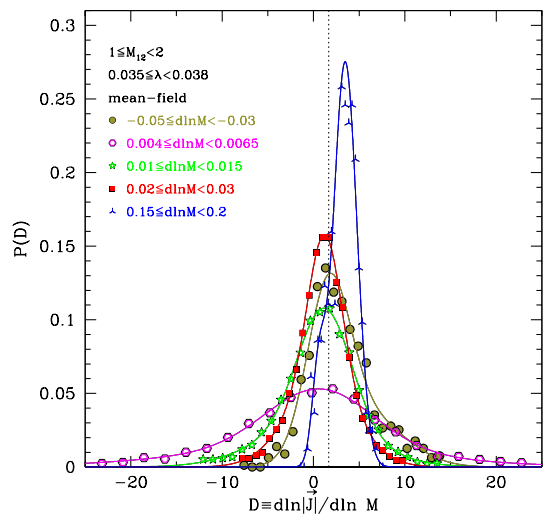}
\caption{
Dependence of $P(D)$ on the amount of mass infall ratio ($\Delta\log_{10}M$)
using samples of $1\le M_{12} <2$, $0.7\le \Delta\rho_{10}<2$, and
$0.035\le \lambda <0.038$. 
}
\label{fig6}
\end{figure}
Figure \ref{fig6} provides an overall view of 
the effect of mass growth ($\Delta\log_{10}M$)
on angular momentum change in samples 
of average spin values ($0.035\le\lambda<0.038$). 
The average value of $D$ (or $\mu_D$) does not deviate much
from $D_c$ (vertical dotted line) 
except in major-merging cases ($0.15 \le \Delta\log_{10}M <0.2$). 
The major merger significantly increases halo spin,
while accretion (or minor-merger) is unlikely to produce a significant 
systematic change in spin for a halo.
The standard deviation in the accretion mode 
seems to be anti-correlated with $\Delta\log_{10} M$, but 
the significance of the anticorrelation is a bit relaxed when weighted with the mass change 
(like $D\times \Delta\log_{10} M$) to update the halo spin (see Eq. \ref{eq:walk}). 

\subsection{Asymptotic Walk of Halo Spin}\label{sec:asym}
\begin{figure}[tp]
\plotone{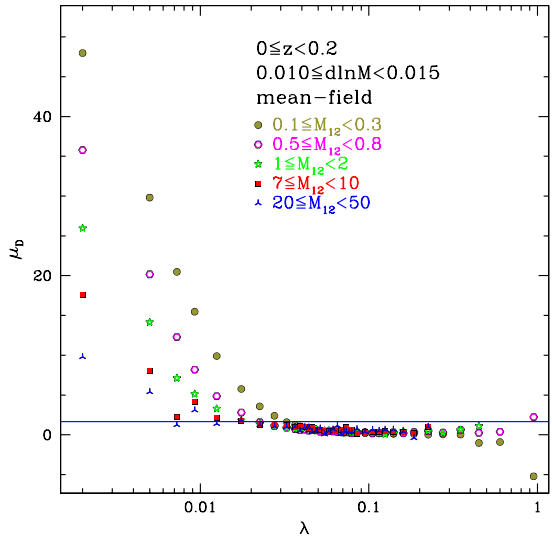}
\caption{
Changes of $\mu_D$ as a function of sample mass
in case of accretion ($0.01\le \Delta\log_{10}M < 0.015$).
The horizontal line is $\mu_D = D_c$. 
}
\label{fig7}
\end{figure}

More interesting features can be obtained if one investigates the dependence of
$\mu_D$ on sample spin.
Figure \ref{fig7} is a typical example showing how  $\mu_D$ changes
with sample mass and spin.
By changing the sample spin while fixing other parameters,
we determine the asymptotic state of the spin distribution.
If $\mu_D>D_{\rm c}$, then the halo spin will increase;
however, if $\mu_D <D_{\rm c}$, then the spin tends to decrease.
If $\mu_D$ has a negative slope and crosses $D_{\rm c}$, then the spin random 
walk (Eq. \ref{eq:walk}) may converge to a characteristic spin value 
where $\mu_D$ crosses $D_{\rm c}$.
If we define $\lambda_{\rm c}$ as a characteristic spin 
as $\mu_D(\lambda_{\rm c}) = D_{\rm c}$,
an the asymptotic state of the spin distribution is produced
with a width depending on the shape of $P(D)$ at $\lambda = \lambda_{\rm c}$.
Virtually every mass sample shows a negative slope, and the range of the characteristic spin 
is $0.01 \lesssim \lambda_{\rm c} \lesssim 0.05$.  
A more massive sample tends to have a lower value of $\lambda_c$.

If $\mu_D < 0$, then the rotational axis ($ \boldsymbol J$) of the halo 
and the direction of the orbital angular momentum ($ \boldsymbol L$) of 
infalling matter tend to be anti-correlated.
However, if $0<\mu_D< D_{\rm c}$, 
then those two vectors show a positive correlation,
while the mass increase is more dominant than the angular momentum increase; 
as a result, the halo spin decreases (see Eq. \ref{spinp}).
If $\mu_D>D_{\rm c}$, then these two vectors show a strong correlation, and 
halo spin tends to increase accordingly. 
Therefore, a positive correlation between two angular momentums does not always
increase the halo spin depending on the amount of angular momentum change.

The behavior of $\mu_D(\lambda)$ may shape a so-called asymptotic state of 
the spin distribution.
If a halo spin is significantly larger or smaller than $\lambda_{\rm c}$,
then the halo may experience a kind of pressure toward $\lambda_{\rm c}$;
as a result, the halo spin tends to move toward $\lambda_{\rm c}$.
Therefore, the value of $\lambda_{\rm c}$ plays an important role in guiding the spin evolution.
Of course, all halos do not immediately jump to $\lambda_{\rm c}$
but show a spread over a certain range of $\lambda$  around $\lambda_{\rm c}$,
which is regulated by the shape of $P(D|\lambda=\lambda_{\rm c})$.
More massive halos tend to have lower $\lambda_{\rm c}$, which means
that lower spin values are preferred in more massive halo samples. This
is consistent with recent findings for mass-dependent spin distributions
\citep{antonuccio-delogu10,knebe08,bett07}.

\begin{figure}
\plotone{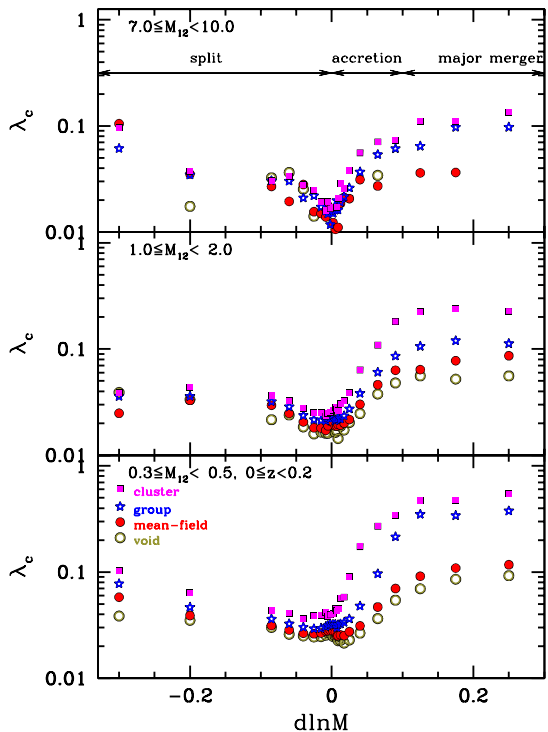}
\caption{
Dependence of $\lambda_c$ on merging mass ($\Delta\log_{10}M$) between $0\le z<0.2$
for three mass samples:
$0.3\le M_{12}<0.5$ (bottom panel), $1\le M_{12}<2$ (middle),
and $7\le M_{12}<10$ (top).
Symbols with different colors are used to distinguish the effect of 
local environment.
}
\label{fig8}
\end{figure}

Figure \ref{fig8} demonstrates how the characteristic spin ($\lambda_c$)
has varied with the infall mass ratio ($\Delta\log_{10}M$) 
in recent epochs ($0\le z <0.2$). 
Broadly speaking,
the halo spin tends to increase as the local density increases and
accretions are less likely to increase the spin value
as much as major mergings at the same local density; meanwhile,
a less massive halo tends to have a larger spin value 
at the same mass increase ($\Delta\log_{10}M$).
In void regions, halos have the lowest
value of $\lambda_{\rm c}$. 
From this description, we conclude that $\mu_D$ in lower
density regions tends to be lower than that in higher density regions 
(or less-aligned infall directions),
and accretion may favor a less biased infall direction lowering the 
characteristic spin. 
Also, smaller cluster halos tend to have more aligned mass infalls 
(or a higher value of $\lambda_{\rm c}$) than larger cluster halos.
The environmental dependence of $\lambda_{\rm c}$ is more clearly observed
in the case of higher $\Delta\log_{10}M$.

\begin{figure}[tp]
\plotone{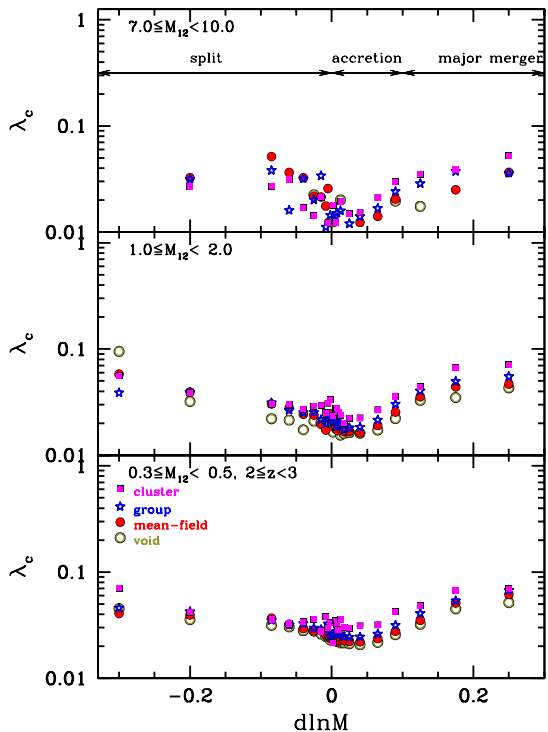}
\caption{
Same as Fig. \ref{fig8} but for redshift samples of $2\le z <3$.
}
\label{fig9}
\end{figure}

The redshift dependence of $\lambda_{\rm c}$ can be observed 
by comparing Figure \ref{fig8} ($0\le  z <0.2$) with Figure \ref{fig9} 
($2\le z<3$).
The dependence of $\lambda_{\rm c}$ on the halo mass at a given local density
is relatively negligible, especially for major merging at high redshifts. 
Overall, $\lambda_{\rm c}$ values at $2\le z <3$ are lower than those for
$0\le z<0.2$ cases, especially for $\Delta\log_{10}M>0$.
The environmental dependence becomes stronger at lower redshifts.

\begin{figure}[ht]
\plotone{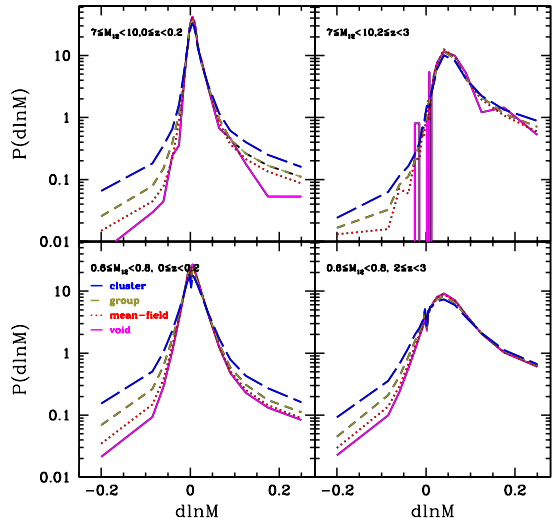}
\caption{
Distribution of $d\log_{10} M$ for halo samples of
$0.6\le M_{12}<0.8$ (bottom panels) and $6\le M_{12}<10$ (top) at 
 $0\le z<0.2$ (left panels) and $2\le z<3$ (right).
}
\label{fig10}
\end{figure}

We consider the frequencies of two infall scenarios (accretion or major 
merging) in Figure \ref{fig10}.
Accretion ($0\le \Delta \log_{10}M <0.1$) at low redshift ($0\le z<0.2$, 
left panels)
is more frequent than at the higher redshift ($2\le z<3$, right panels).
Moreover, halos in less dense regions at low redshifts
are likely to have less frequent major-merging events; 
this environmental dependence is not observed at high redshifts.
Therefore, we conclude that major-merging events subside with time, and 
this tendency becomes stronger in lower density regions. This may be due to the
accelerating expansion of the universe at lower redshifts. 
The acceleration may reduce the number of major-merging events; 
therefore, accretion would be a dominant source of mass increase in recent 
epochs.
Lower-density regions, especially voids, may suffer more seriously from such
acceleration.

\subsection{Random Walk and Correlated Mass Infall}
We need to clarify two confusing terms: random walk and 
correlated (or biased) infall.
In this study, a random walk means that the angular momentum change at a given merging step
is purely stochastic, obeying the distribution of $D$.
However, the $D$ distribution (or $P(D)$) has, in most cases, has
a finite mean value:
a positive bias ($\mu_D>0$) or negative bias ($\mu_D<0$).
The correlated infall direction indicates a preferential direction for
 mass infall 
with respect to the halo rotation axis (or its spin axis; 
see Appendix \ref{appen1} for details) and
does not conflict with the random-walk hypothesis.

Since $\mu_D$ is finite (and mostly positive),
the orientations of the orbital angular momentum of infalling matter 
are not perfectly random but are somewhat correlated or anti-correlated
with the halo rotational axis (or the direction of the angular momentum vector).
However, it should be noted that as the halo spin approaches zero, even a small amount of mass infall 
(or accretion) may significantly increase the halo spin value 
($|{\boldsymbol L}| \gg |{\boldsymbol J}| \simeq 0$),
which may explain the sharp increase in $\mu_D$ as the sample spin 
approaches zero (Fig. \ref{fig7}).

An illustration of a spin random walk is a staggering walk taken by a 
drunken man. 
Consider a half-pipe with a U-shaped cross section and, then,  a group of 
drunken men take a walk lurching forward along the bottom of the half-pipe 
without mutual collisions. 
Here, the shape of the cross section is not fixed, and the pipe is
not like $P(D)$, which varies with sample parameters.
Therefore, the distribution of walker positions gradually changes to
adjust itself to the cross-sectional shape of the half-pipe, and
the majority of walkers are distributed near the bottom of the pipe ($\lambda_c$ in the halo spin walk).
In the next section, we investigate the results of spin random walks
and the effects of various parameters on the shape of spin distributions.

\section{Random Evolution of Halo Spin}\label{sec:spin}
In this section, we will show how well the stochastic model reproduces  distributions of 
halo spin measured from $N$-body simulations. The change in angular momentum is expected to 
be a random process with a distribution that has a bimodal Gaussian form.
We assume that the spin change in each time step is independent of the
previous history (a Markov process; 
see  \citealt{bond91} for an example applied to halo mass function)
and is completely determined by the current physical status.
However, it is possible that subsequent 
infall events can be correlated with previous events.
For example, consider successive mass infalls through a local filamentary structure. 
In this case, the random-walk hypothesis is not satisfied. 
We will discuss possible correlated infalls in the latter part of this section.

\subsection{Stochastic Model of Spin Evolution}
First, we choose a mass history from simulated 
main merger trees. The initial spin value is set to $\lambda_0=0.01$.
At each merging step, we identify $P(D)$ of a subsample with parameter
ranges enclosing the current halo mass, spin, mass change, local density, 
and redshift.
Then, from the identified $P(D)$ we randomly 
generate a value of $D_{\rm rand}$.
From equation (\ref{eq:walk}) with $D_{\rm rand}$, 
we calculate the change in spin ($\Delta\lambda$) and update the spin value. 
In the next merging step, we iterate the same process but with the
updated spin value. 
The resulting spin value ($\lambda_f$)
at the final redshift is obtained as
\begin{equation}
\lambda_f = \lambda_0 + \sum_{i=1}^N \Delta \lambda_i.
\end{equation}

\subsection{Effect of Halo Mass on Spin Distribution}
Now, we obtain the 
spin distribution produced by the stochastic random walk
and compare it with that of $N$-body simulations.
\begin{figure*}[tp]
\plotone{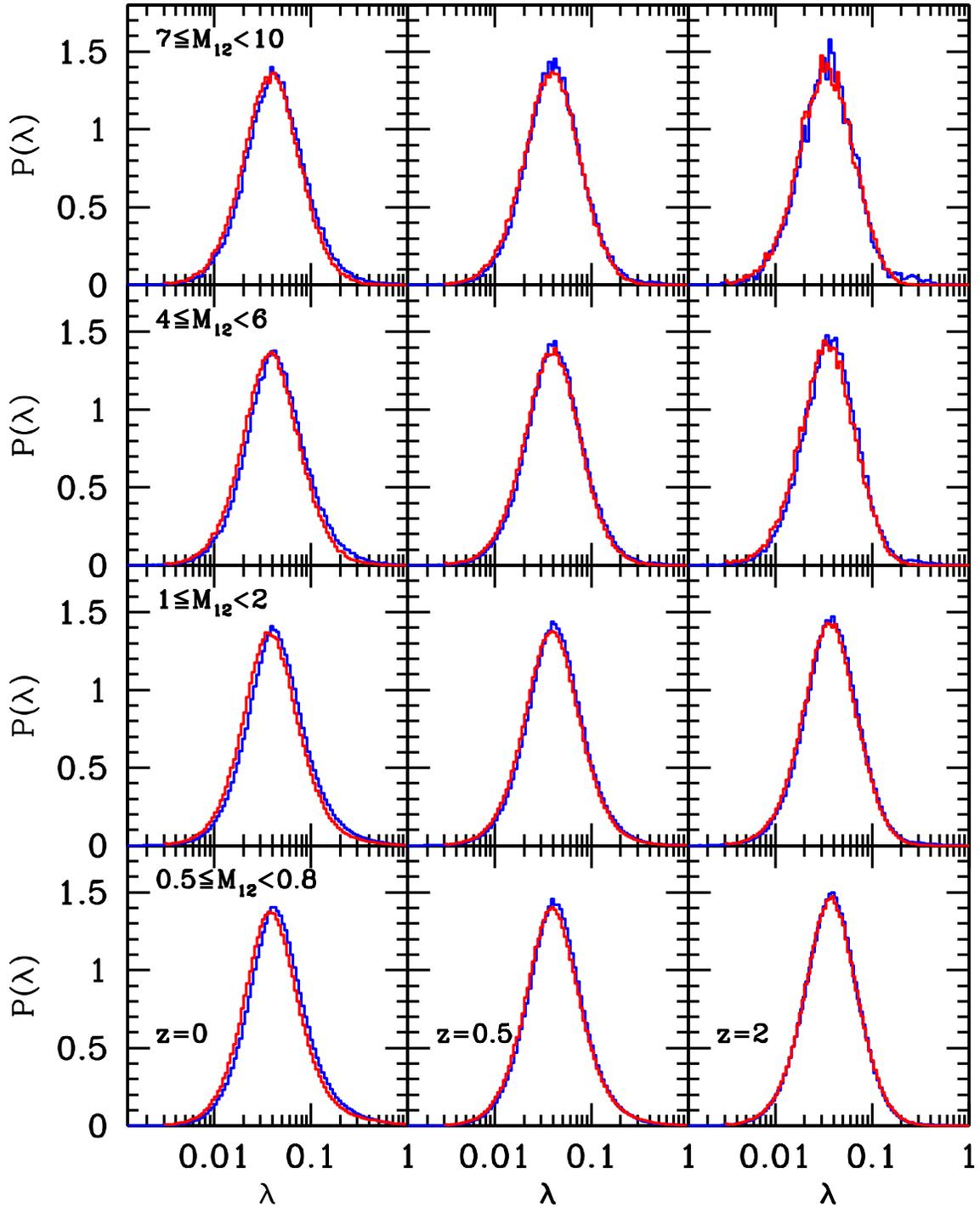}
\caption{
Spin distributions of various halo mass
samples at $z=0$, 0.5, and 2 (from left panels).
The blue and red histograms are the randomly 
generated and $N$-body-simulated spin distributions, respectively.
}
\label{fig11}
\end{figure*}
Figure \ref{fig11} shows the results with randomly generated spin distributions (blue histogram)
against $N$-body simulated distributions (red histogram)
in four mass samples.
The randomly generated spin distributions at $z=0$ are similar to the 
simulated distributions with a slight shift to higher $\lambda$, 
and this difference is observed in every mass sample,
although the difference is negligible at higher redshift.

\begin{figure}[tp]
\plotone{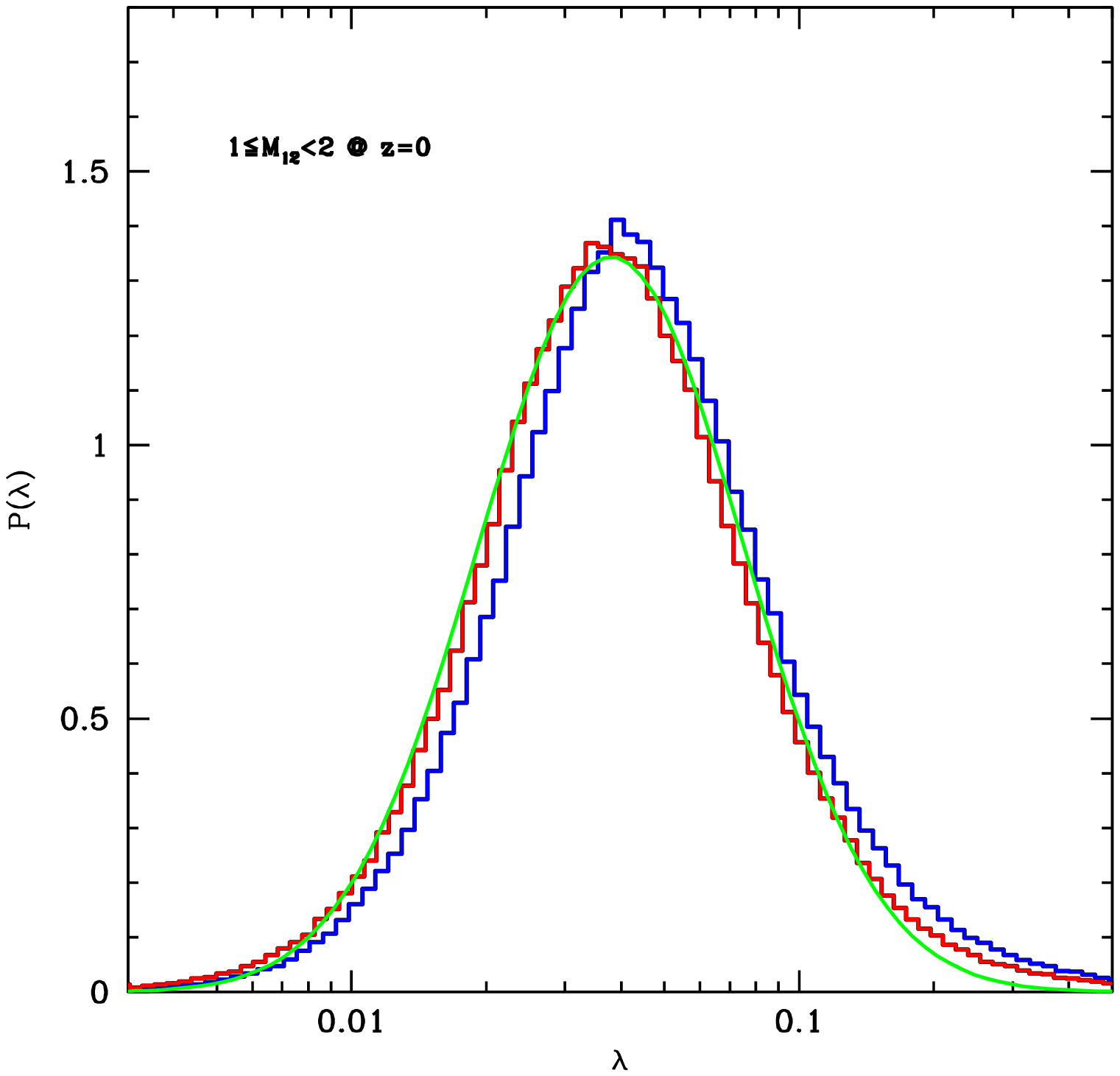}
\caption{
Spin distributions of the mass sample of $1\le M_{12}<2$ at $z=0$.
Blue and red coded histograms are the 
randomly generated and $N$-body spin distributions, respectively, 
and the green curve shows the log-normal fit to the $N$-body histogram.
}
\label{fig12}
\end{figure}

\begin{figure}[tp]
\plotone{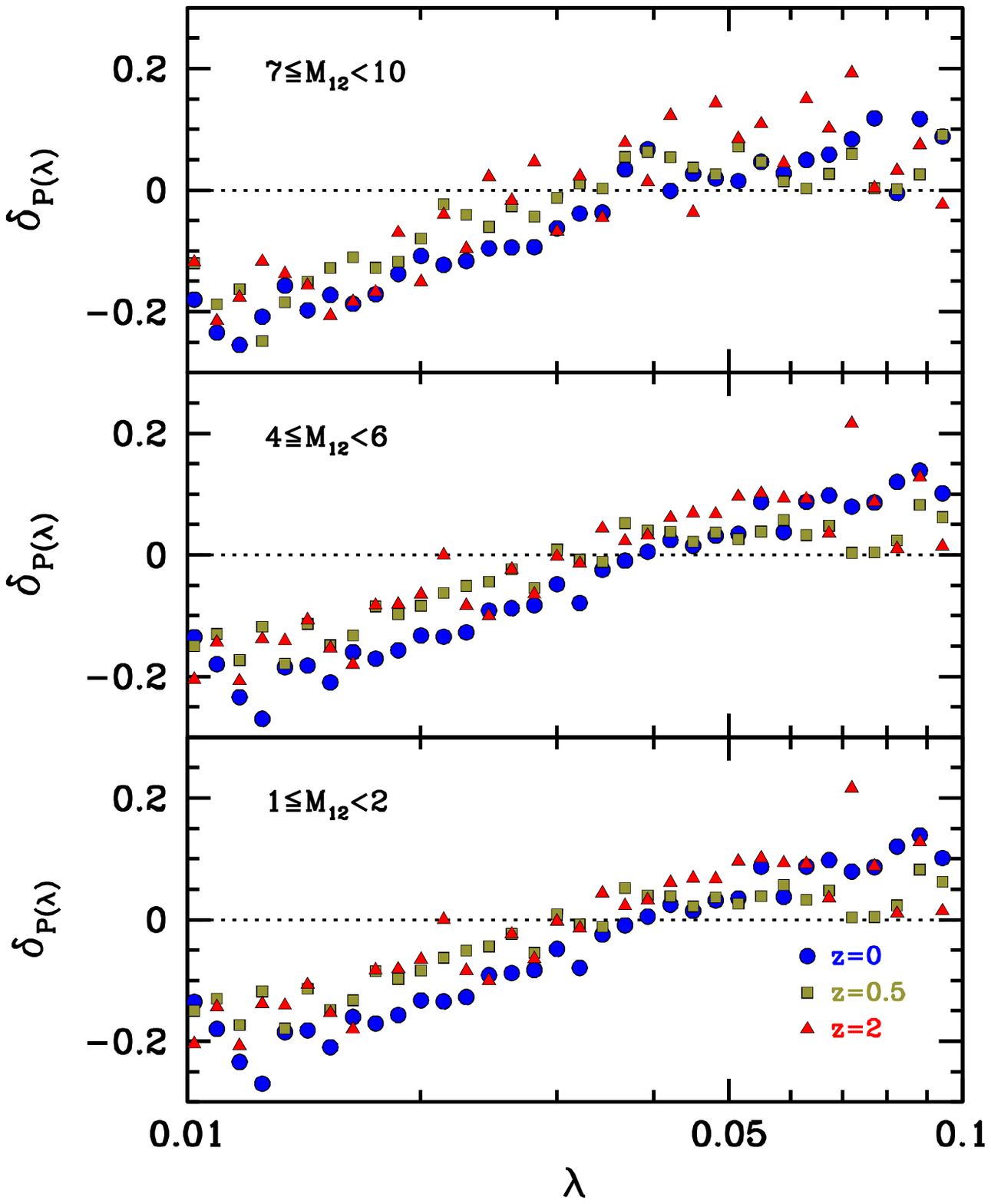}
\caption{
Deviations in spin distribution for three mass samples measured in the spin interval
of $0.01\le\lambda<0.1$. 
The circles, boxes, and triangles
are measured at $z=0$, 0.5, and 2, respectively. 
We removed the irrelevant regions of $\lambda<0.01$ and $\lambda>0.1$
as the number of simulated halos in those regions is very small.
}
\label{fig13}
\end{figure}

In Table~\ref{tab:fit1} in Appendix~\ref{appen5}, 
the log-normal fitting results for randomly generated and
simulated spin distributions are given with a mean value, standard deviation,
and $\chi^2$/degrees of freedom (dof).
This difference is more clearly shown 
in Figure \ref{fig12}, where the blue and red histograms are the distributions 
of the random and simulated spins at $z=0$, respectively.
The green curve is a Gaussian fit to the simulated spin distribution.
Before jumping to a detailed comparison, it would be valuable to note that
$N$-body-simulated halos overpopulate at the high-$\lambda$ tail 
(the same feature is also observed for the generated spin distributions)
compared to the log-normal fitting (solid curve), 
which has also been found in other studies \citep{bett07,shaw06},
unless a correction is applied to the halo
samples \citep{davis11,shaw06}. 
In Figure \ref{fig13},
we show the difference ($\delta_{P(\lambda)} \equiv (P_r-P_s)/P_s$)
between the simulated ($P_{\rm s}$) and randomly generated ($P_{\rm r}$) 
spin distributions.
The differences increase as redshift decreases,
while deviations in the randomly generated spin distributions at $z=0.5$ and 2
are confined within about 10--15\% error.


\subsection{Environmental Effect on Spin Distribution}\label{sec:env}
If a spin walk is well described by the Markov (or random) process and a sufficient 
number of trials are performed to suppress the Poisson error, then
the $N$-body-simulated and randomly generated
spin distributions should be basically equal to each other.
Therefore, a subtle failure in recovering the simulated spin distribution
at lower redshifts, as seen in Figure \ref{fig11}, may imply a possible
limitation of the stochastic model.

A possible explanation for the mismatch of spin distributions
between the simulation and the stochastic model
is correlated mass merging, which means
a subsequent mass infall changes the angular momentum of a halo in the same direction.
Then, a set of $D_{\rm rand}$s in the successive infalls may not be random but correlated.
This correlated mass merging could be a function of local environment.
In void regions, halos are likely to have fewer major mergers, while those 
in cluster regions may experience more frequent heavy mass infalls (see Fig. \ref{fig10}).
Because it is believed that major mergers frequently occur when clumps of 
matter fall along filamentary structures,
the number of filaments connected to a halo determines whether or not
the spin walk is random. In void regions, filamentary structures are 
relatively rare and void halos tend to have a small number of filaments around them,
which means that subsequent major mergers may occur along a smaller 
number of filamentary structures.

\begin{figure}[tp]
\plotone{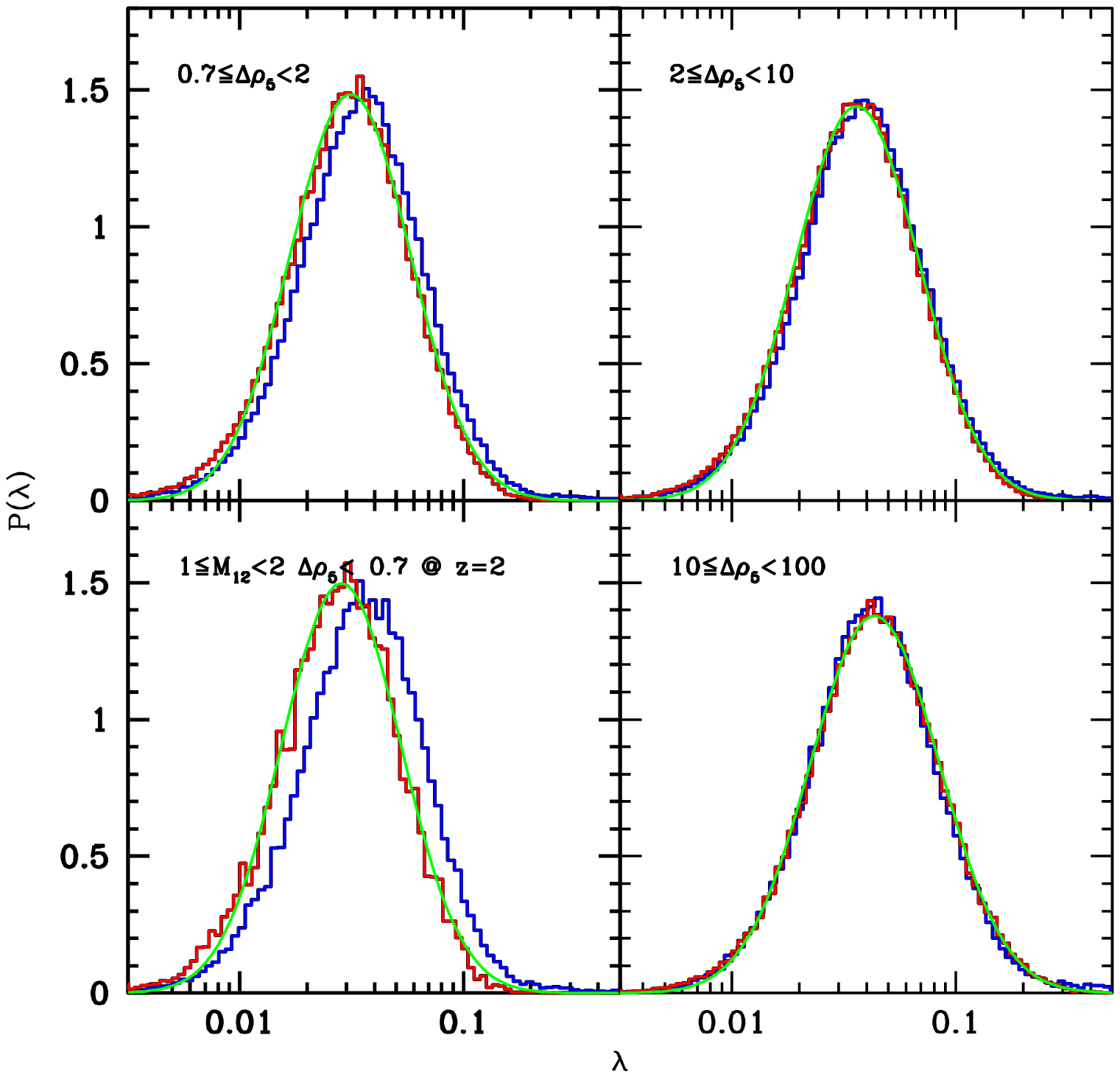}
\caption{
Spin distributions in various local environments at $z=2$.
Counterclockwise from the bottom left panel are the
spin distributions of $N$-body (red histogram) and random-generated (blue) samples of
local densities of $\Delta\rho_{10}<0.7$, $0.7\le\Delta\rho_{10}<2$, $2\le\Delta\rho_{10}<10$,
and $10\le\Delta\rho_{10}< 100$.
The green solid curve in each panel is a log-normal fit to the $N$-body spin distribution.
}
\label{fig14}
\end{figure}

Figures \ref{fig14}--\ref{fig16} show the environmental 
effect on spin distribution at three redshifts.
In Table~\ref{tab:fit2} of Appendix~\ref{appen5}, 
the log-normal fitting results are listed.
We clearly see that a randomly 
generated spin (blue histogram) tends to substantially
overestimate the halo spin of the $N$-body-simulation in lower density regions.

For the group and cluster samples,
the difference in spin distributions between the simulation and 
the stochastic model is relatively small.  
However, with time the difference increases with time
and the group sample begins 
to show such a discrepancy at $z=0$.

\begin{figure}[tp]
\plotone{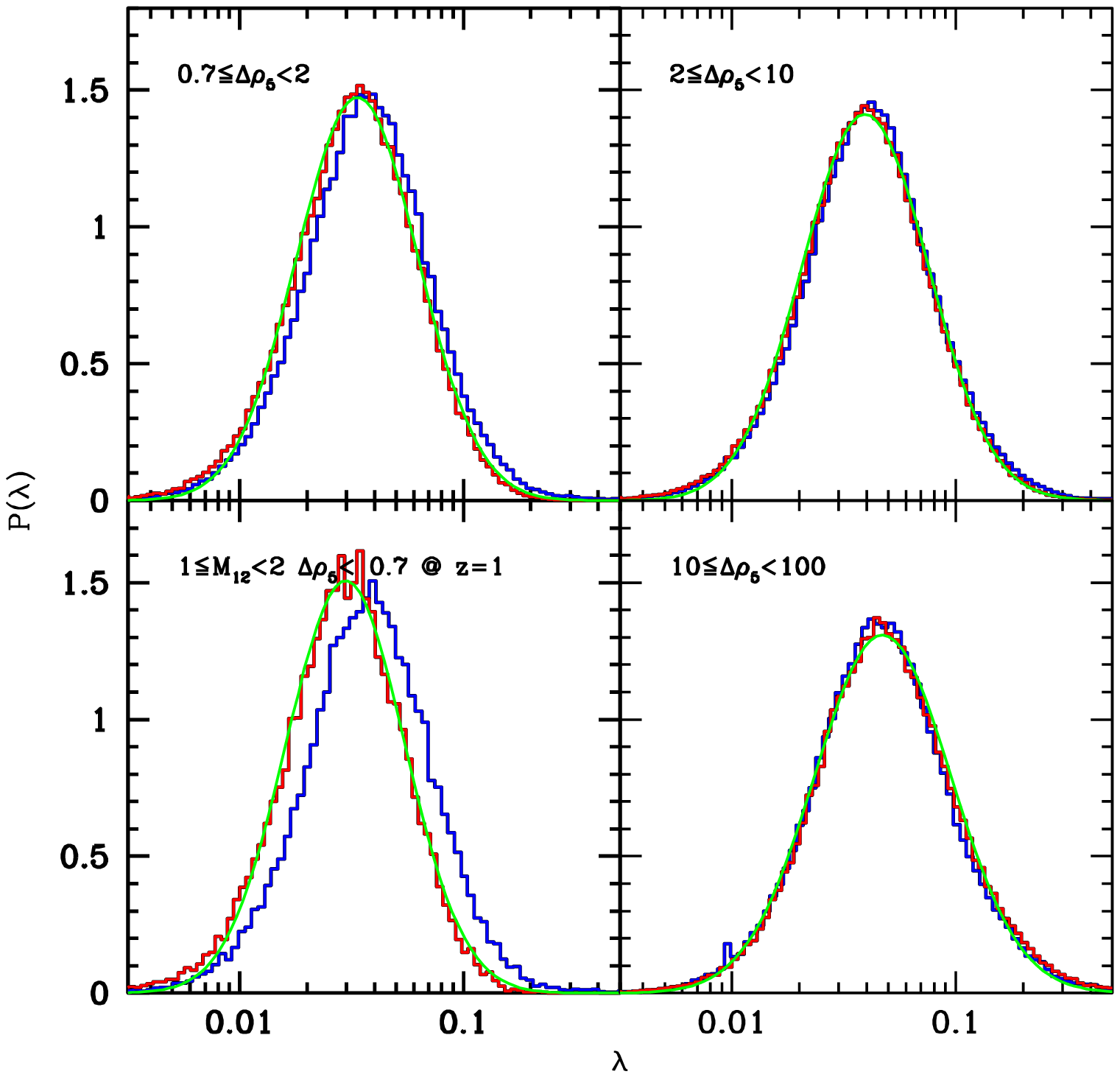}
\caption{
Same as Fig. \ref{fig14} but at $z=1$.
}
\label{fig15}
\end{figure}
\begin{figure}[tp]
\plotone{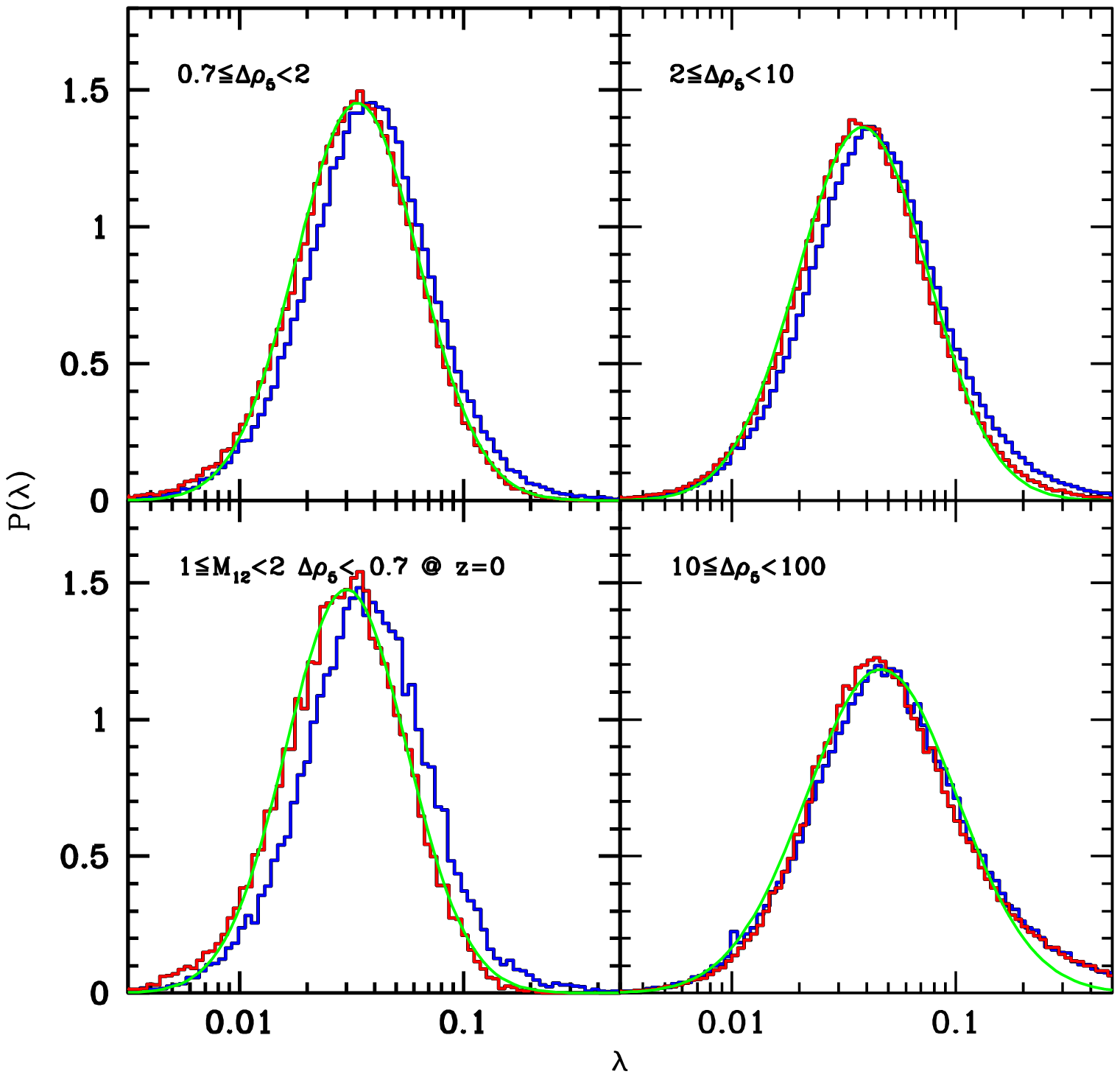}
\caption{
Same as Fig. \ref{fig14} but at $z=0$.
}
\label{fig16}
\end{figure}

\begin{figure}[tp]
\plotone{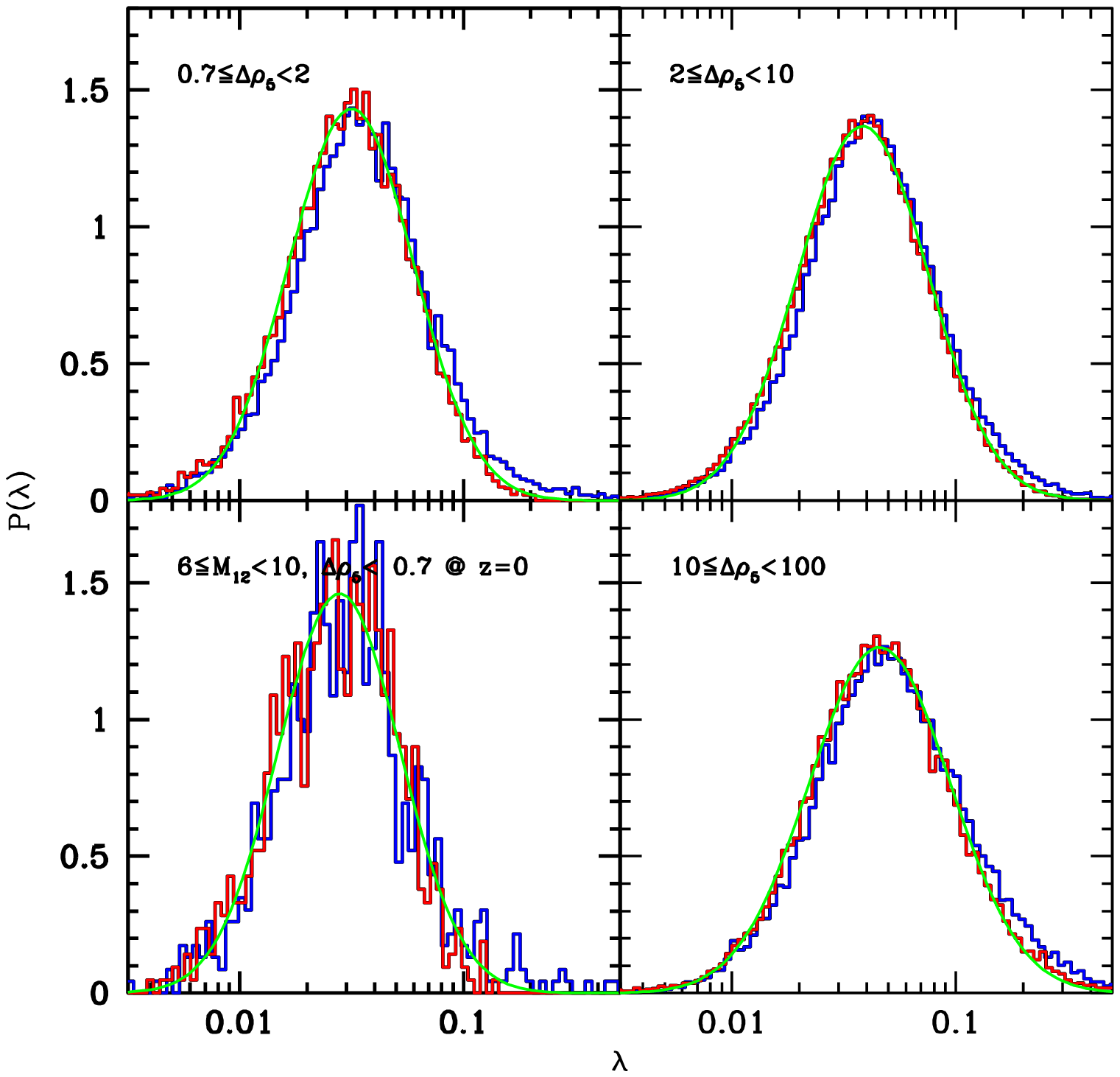}
\caption{
Spin distributions of a more massive sample of $6\le M_{12}<10$ at $z=0$.
The green solid curve is a log-normal fit to the corresponding $N$-body distribution.
}
\label{fig17}
\end{figure}

Since a massive halo is likely to be located in a knot (interconnection) of local 
filamentary structures, the probability of successive major mergers along a single filament
is low (the next major merger may occur with another filamentary structure).
Therefore,
it would be interesting to determine 
whether the randomly generated halo spins of more massive samples
have a distribution similar to that of the $N$-body simulation.
In Figure \ref{fig17}, we show the generated spin distributions of
a massive sample of $6\le M_{12}<10$ in different local environments.
No substantial difference is shown even in void regions.
Therefore, it is quite reasonable to believe that massive halos tend to
show random evolution in spin. 

In these figures, 
the halo spin distribution depends on the local environment;
denser regions cause halos to rotate faster, which is also confirmed 
in Figure~\ref{fig8} ($\lambda_{\rm c}$ is higher in a denser region).
Also, most of the spin samples show log-normal distributions, except for the 
cluster samples at $z=0$, where the distribution has a slightly fat tail 
at larger $\lambda$.
This may be due to unrelaxed halos in cluster regions where violent mergers
are so frequent that cluster halos do not have sufficient time to transfer 
angular momentum to outside
and finally to be virialized.

\section{Markovian and Non-Markovian Processes}\label{sec:markov}
Throughout this paper, the stochastic model we adopted assumes 
that the spin walk can be described by the Markov process, 
which requires that a random walk be memoryless or independent of previous 
states. 
Therefore, it is important to determine whether the spin walk observed 
in $N$-body simulations is really random (Markovian) or not (non-Markovian).

\subsection{Autocorrelation between Neighboring Mass Events}
One of the easiest ways to verify the correlation of two separate events in 
a halo merging history is to measure 
the autocorrelation (or serial correlation) between a pair of events in a main merger tree line.
The first-order autocorrelation of two mass events of $i$ and $j$ is
\begin{equation}
\Xi(i,j) = {{ w_iw_j\left({D_i-\mu_i \over \sigma_i}\right) \left({D_j-\mu_j\over \sigma_j}\right)} }\end{equation}
where $w$ is a weight, $D$ is randomly generated by the stochastic model
or provided by simulations,
$\mu$ is the mean, and $\sigma$ is the standard deviation of the probability
distribution of $D$. In this analysis, we adopt $w_i = |\Delta\log_{10} M_i|$
and define the lag-$k$ autocorrelation as
\begin{equation}
C_k(z) = {1\over {\mathcal{W}}_{\rm pair}} \sum_{i,j=i+k} {\Xi(i,j| \ge z)},
\label{eq:auto}
\end{equation}
where ${\mathcal{W}}_{\rm pair} = \sum w_i w_j$ and
the summation is performed over all possible lag-$k$ pairs of events
in a halo history to redshift $z$.
Only a relative comparison is possible because the time interval between a lag-$k$ pair changes.

\begin{figure}[tp]
\plotone{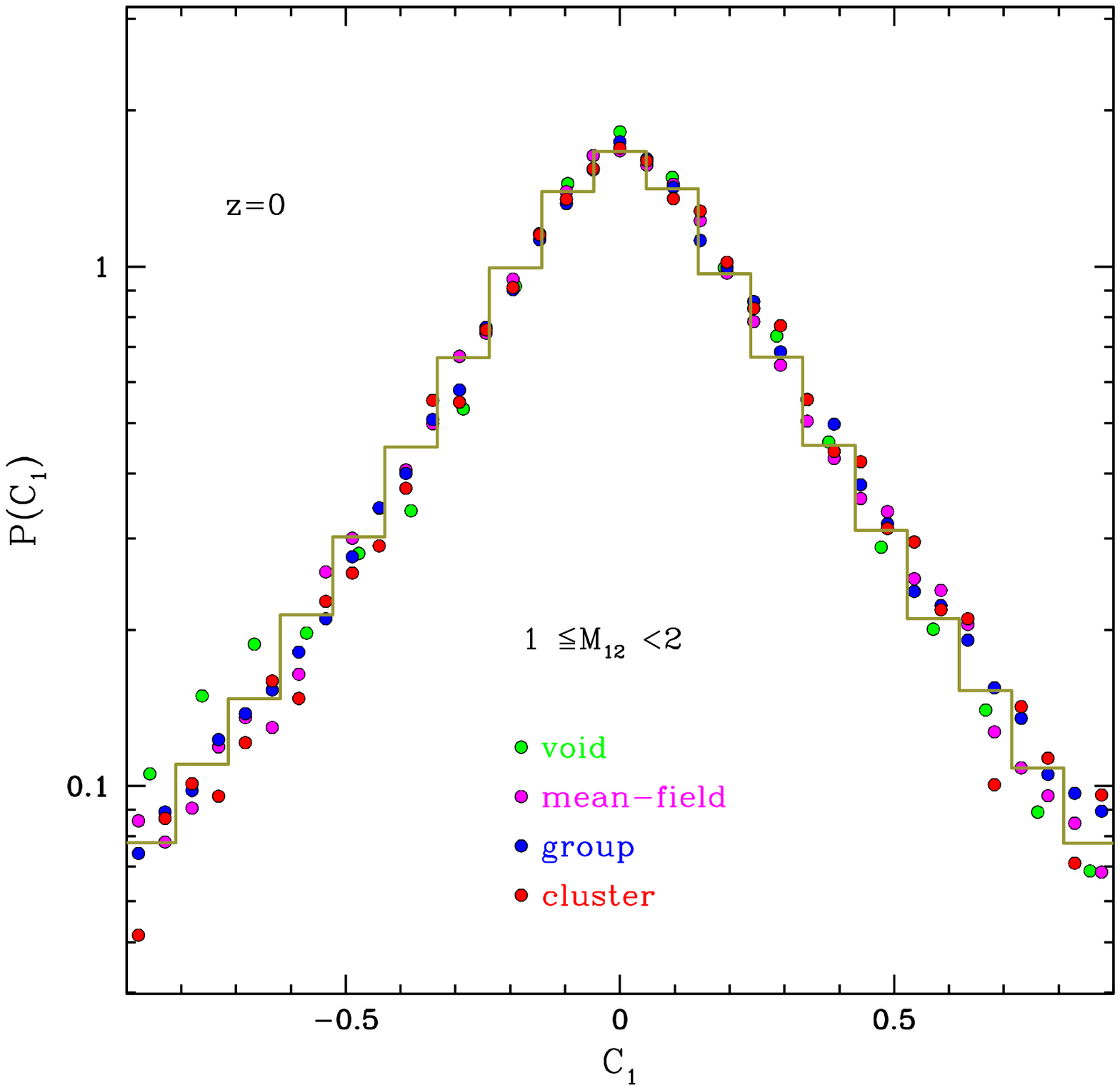}
\caption{
Example of the probability distribution of $C_1$ (Eq.~\ref{eq:auto}) 
in a sample of $1\le M_{12}<2$ at $z=0$.
The symbols are distributions measured with the spin history of the
$N$-body halo sample 
with $\Delta\rho_{10}<0.7$ (green), $0.7\le\Delta\rho_{10}<2$ (magenta), and $2\le\Delta\rho_{10}<10$ (blue). 
The histogram shows the mean distribution of 100 random generations in the sample of $\Delta\rho_{10}<0.7$.
}
\label{fig18}
\end{figure}
Using 100 random realizations of $P(C_k)$,
we statistically measured
the deviation of the autocorrelations of $N$-body data from random expectations.
The probability distribution of a lag-one autocorrelation is shown in Figure \ref{fig18}
for halo samples of $1\le M_{12}<2$
in four local-density regions (colored symbols).
The brown histogram is a mean expectation measured 
from the 100 random realizations. 
Even though the probability distribution of $D$ depends on the local density,
the autocorrelation seems to be indistinguishable between different local-density samples
and even from the random model in this plot.

\begin{figure*}[tp]
\plotone{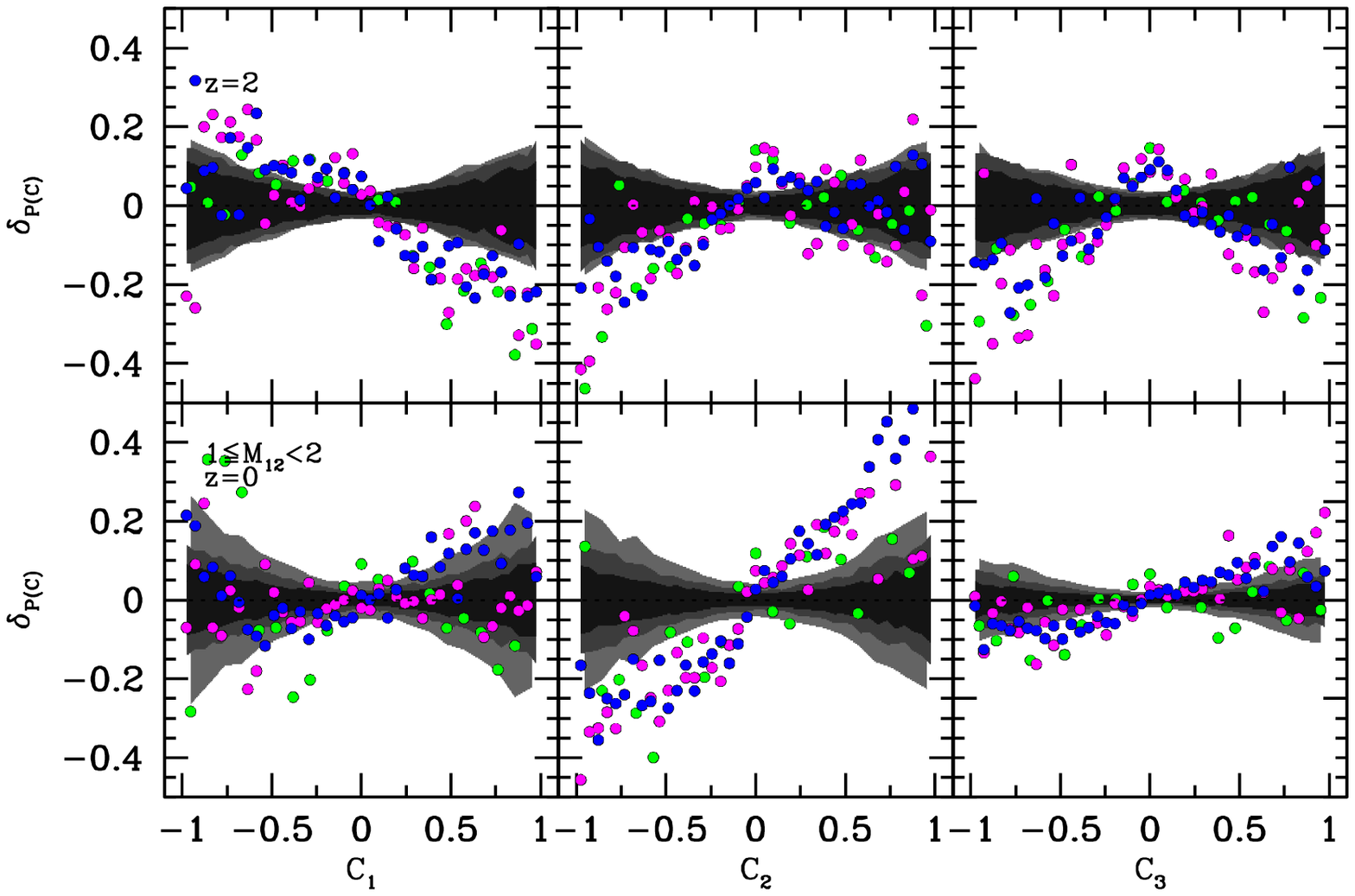}
\caption{
Randomness check by the measurement of autocorrelations, $C_1$ (left panels), $C_2$ (middle), and $C_3$ (right)
in the sample of $1\le M_{12} <2$ at $z=0$ (bottom panels) and 2 (top).
The three gray levels in the regions are the 1--$\sigma$ distributions
of 100 random generations for $\Delta\rho_{10}<0.7$ (outer gray region),
$0.7\le\Delta\rho_{10}<2$ (middle), and $2\le\Delta\rho_{10}<10$ (innermost)
with corresponding $N$-body estimations plotted with green, magenta, and blue circles, respectively.
}
\label{fig19}
\end{figure*}

\begin{figure*}[tp]
\plotone{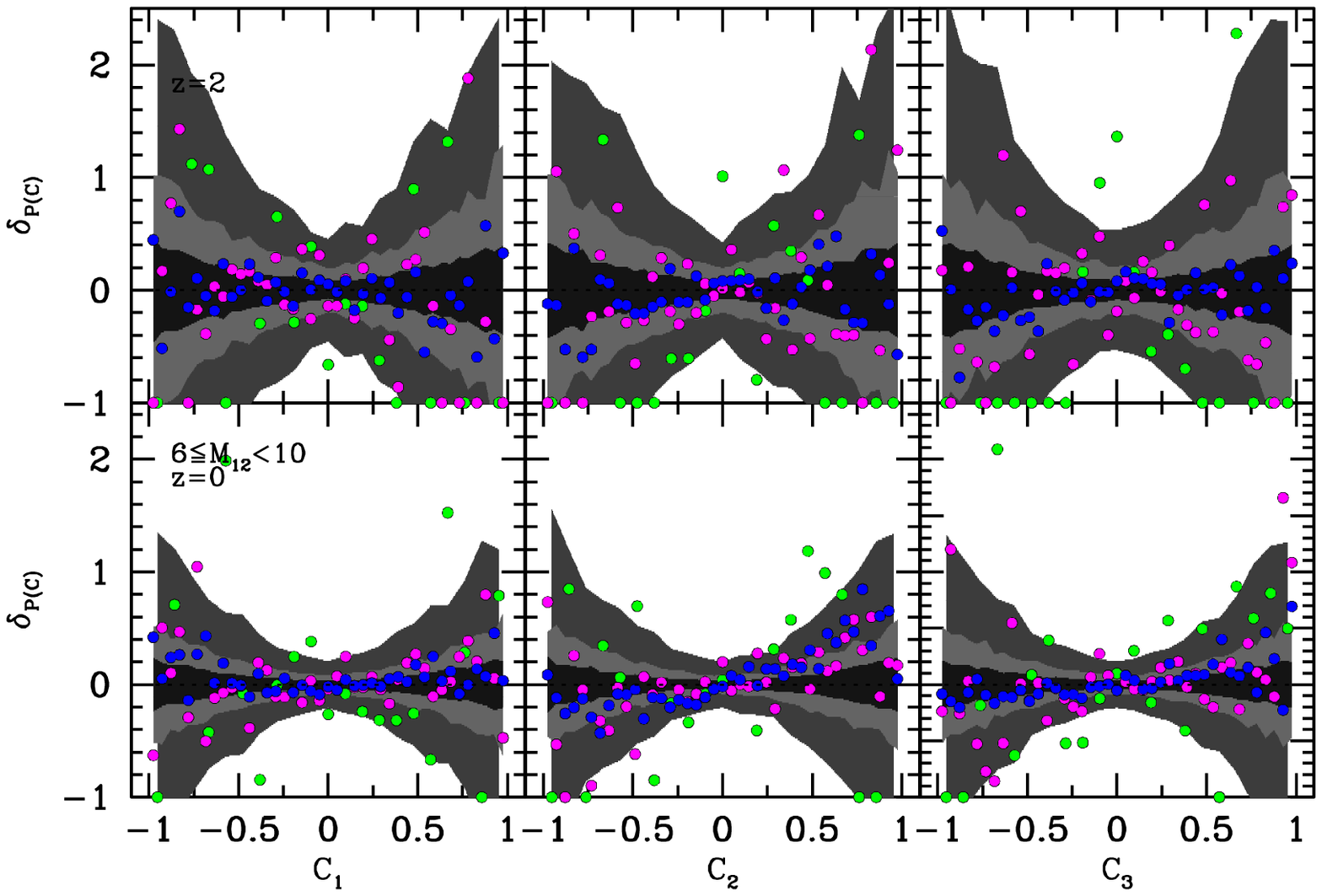}
\caption{
Same as Fig. \ref{fig19} but for the massive samples of $6\le M_{12}<10$.
Due to the smaller sample size, the 1--$\sigma$ scatter of the random sample is larger than that of the sample of $1\le M_{12}<2$.
}
\label{fig20}
\end{figure*}

We also showed deviations between simulated correlations and the random model
with $\delta_{P(C_k)} \equiv (P_s(C_k)-P_{<r>}(C_k))/P_{<r>}(C_k)$,
where $P_s(C_k)$ is the simulated autocorrelation and $P_{<r>}(C_k)$ 
is the average of the 100 randomly generated autocorrelations.
Figure \ref{fig19} shows the results
for a sample of $1\le M_{12}<2$.
The shaded region depicts the 1$\sigma$ distributions of the random model,
and the symbols are measured from $N$-body simulations.
It is easy to see the negative autocorrelations of $C_1$ at $z=2$ and
the positive correlations of $C_2$ at $z=0$
with greater than 2$\sigma$ confidence limits.
We further investigated autocorrelations from $C_4$ to $C_6$ 
but found nothing substantially deviating from the random model.
On the other hand, we could not detect any possible autocorrelations in the 
massive sample ($6\le M_{12}<10)$). Most of the 
simulated autocorrelations are contained
within 1$\sigma$ random expectations (see Fig. \ref{fig20}).

Detecting substantial autocorrelations of $C_1$ at $z=2$ and $C_2$ at $z=0$ may provide
a tangible clue for void discrepancies 
between the simulated and generated spin distributions.
Correlated infall events may cause void halos to have smaller spin values,
but we are unable to detect any environmental dependence in the correlation 
figure. However, considering the frequent
major merging of cluster halos, an autocorrelation could easily be 
counterbalanced by other major-merging events. 
However, the void halos have little possibility of another major merging.

\subsection{Origin of the Log-normal Distribution of Halo Spin}
In this subsection, we demonstrate that the log-normal distribution is 
a natural consequence of the stochastic differential equation in the 
stochastic of halo spin.
In many studies, simulated spin distributions in various mass samples 
consistently show substantial deviations from the log-normal form in
the high-$\lambda$ tail.
However, no serious deviations are found
as long as a mass sample is further divided into local-density subsamples.
As shown in Figures \ref{fig14}--\ref{fig16},
the log-normal fitting functions (green curves) seem to describe well the
simulated spin distributions (red histograms) in various local environments,
except for cluster regions at $z=0$.

Now, consider a one-factor model of the Markov process 
or the model of Geometric Brownian Motion (GBM; \citealt{ross10}), 
which has a stochastic differential equation of
\begin{equation}
{d\log_{10} \lambda(\tau) \over d\tau} = \theta + \sigma_c {dW_\tau\over d\tau},
\label{gbm}
\end{equation}
where $\theta$ describes the long-term drift of the system, $\sigma_{\rm c}$ 
is a constant, 
and $W_\tau$ is a kind of normally distributed Wiener process or 
$W_\tau \sim {\mathcal{N}} (0,\tau)$
where ${\mathcal{N}}(0,\tau)$ denotes
the normal distribution with a zero mean value and a variance of $\tau$.

From the properties of the Wiener process such as $(W_{\tau_2}-W_{\tau_1}) 
\sim {\mathcal{N}}(0,{\tau_2}-{\tau_1})$
for $\tau_2\ge \tau_1$, 
we obtain the probability distribution of $dW/d\tau$ as
\begin{equation}
{dW_\tau\over d\tau} = { W_{\tau+d\tau} - W_\tau \over d \tau} \sim {{\mathcal{N}}(0,d\tau)\over d\tau}
= {\mathcal{N}} (0, 1/d\tau),
\label{wienereq}
\end{equation}
where the statistical relation of 
$\mathcal{N}(\mu,\sigma^2)/a = \mathcal{N}(\mu,\sigma^2/a^2)$ is used.
Thus, the probability distribution of $dW_{\tau}/d\tau$ is 
a normal distribution with a zero mean and a variance of $1/{d\tau}$.
The Wiener process has the following properties:
\begin{eqnarray}
dW_\tau &\sim& \mathcal{N}(0,1) \sqrt{d\tau}\\
dW^2_\tau &\sim& \sqrt{2}\mathcal{N}(0,1) d\tau \propto d\tau.
\label{weq}
\end{eqnarray}

According to It\={o}'s formula \citep{movellan11}, we may linearize the change of $\log \lambda$ as
\begin{eqnarray}
d\log \lambda &=& {d\log\lambda \over d\lambda} d\lambda + {1\over2} {d^2\log \lambda \over d\lambda^2} d\lambda^2\\
&=& \sigma_c dW_\tau + \left({ \theta - {\sigma_c^2\over 2} }\right) d\tau
\end{eqnarray}
where we used the relation $d\lambda = \lambda (\theta d\tau + \sigma_c dW_\tau)$ 
(from Eq.~\ref{gbm})
and $dW{_\tau}^2 = d\tau$ (Eq.~\ref{weq}), and ignored
the higher-order terms of $d\tau$.
By integrating the above equation, we finally obtain
\begin{equation}
\log\left({\lambda(\tau) \over \lambda_0}\right) = {\sigma_c W_\tau + \left({\theta - {\sigma_c^2\over 2}}\right) \tau }.
\end{equation}
Now, one can understand that 
the logarithm of the stochastic process $\lambda$ follows a Wiener process
with a standard deviation of $\sigma_c$
and a corrected long-term drift of $\theta - \sigma_c^2/2$
(\citet{oksendal} for details).
If $W_\tau$ does not have a normal distribution, then
we may not obtain a logarithm of the randomly distributing quantity.
For example, 
if $W_\tau$ has a probability proportional to $-\tau$ or $-\tau^2$, then
the resulting distribution would be exponential or Gaussian, respectively.

We now attempt to combine the right two right-hand terms in equation (\ref{gbm}) 
into $\mathcal{D}$ as
\begin{equation}
{d\log_{10} \lambda(\tau)\over d\tau} = \mathcal{D},
\label{lognormal1}
\end{equation}
where $\mathcal{D}\sim \mathcal{N}[\theta, \sigma_c^2/{d\tau}]$.
After changing variables as $\tau\rightarrow\log_{10} M$ 
and $\mathcal{D} \rightarrow D-D_c$,
we can find an analogy between the GBM and 
the angular momentum change, $D$ (see Eq.~\ref{eq:walk}).
Although the overall distribution of $D$ is better fit by the bimodal 
Gaussian, the distribution around a peak can be well modeled by a single 
Gaussian peak because the two Gaussian components have nearly the same center 
in most samples, and 
we may simply approximate $P(D)$ around the peak as a normal distribution.
The minor Gaussian component with a broad width is mostly needed to describe 
the fat tails on both sides.
Consequently, the distribution of $\log_{10} \lambda$ follows the normal 
process, which means that the spin distribution of $\lambda$ should accordingly 
be log-normal.
If the centers of two Gaussian components are well separated, or the minor 
Gaussian component is comparable to the major one, then 
the resulting distribution 
of $\log_{10}\lambda$ may show a considerable deviation from the normal 
distribution.

Here, we need to address the validity of the application of the GBM model 
to spin evolution before jumping to a conclusion, and thus
we test whether the halo spin change satisfies two prerequisites 
of the GBM model:
\begin{itemize}
\item  $\theta$ (or $\mu_D$) should depend only on $\tau$ (or $\log_{10}M$); and
\item 
the standard deviation of $\mathcal{D}$ (or $\sigma_D$) should be proportional to 
$d\tau^{-1/2}$ (or $(\Delta\log_{10} M)^{-1/2}$).
\end{itemize}
To qualitatively assess the first condition, 
Figures \ref{fig3}, \ref{fig4}, \ref{fig5}, and \ref{fig6}
show that there is no significant change in $\mu_D$ with different redshifts, 
local environments, or infall mass ratios, except for major-merging events, in 
which case a substantially larger $\mu_D$ is obtained.
Consequently, one can easily see the mass dependence in Figure \ref{fig7}.

The second condition might be related to the temporal resolution affecting the mass infall ratio.
Now, we investigate the following relation:
\begin{equation}
 \sigma_{{D}}^{-1} =   (\Delta \log_{10} M)^{1/2}  \sigma_c^{-1},
\end{equation}
where $\sigma_c$ is constant with respect to $\Delta \log_{10} M$ 
as long as the second requirement is satisfied.
In order to obtain a log-normal distribution of $\lambda$, 
$\sigma_{{D}}^{-1}$ should be linearly scaled to the change in halo mass.
As shown previously, the spin distribution has 
a mass and environmental dependences, 
indicating that the distribution is not stationary 
but evolves with the halo mass.
However, halo environments do not change substantially over cosmic time
since a halo does not generally move a great distance ($\lesssim$ a few Mpc).
Hence, it is sufficient to examine the relation between $\sigma_{{D}}^{-1}$ 
and $\Delta \log_{10}M$ in the same environment.

\begin{figure}[tp]
\plotone{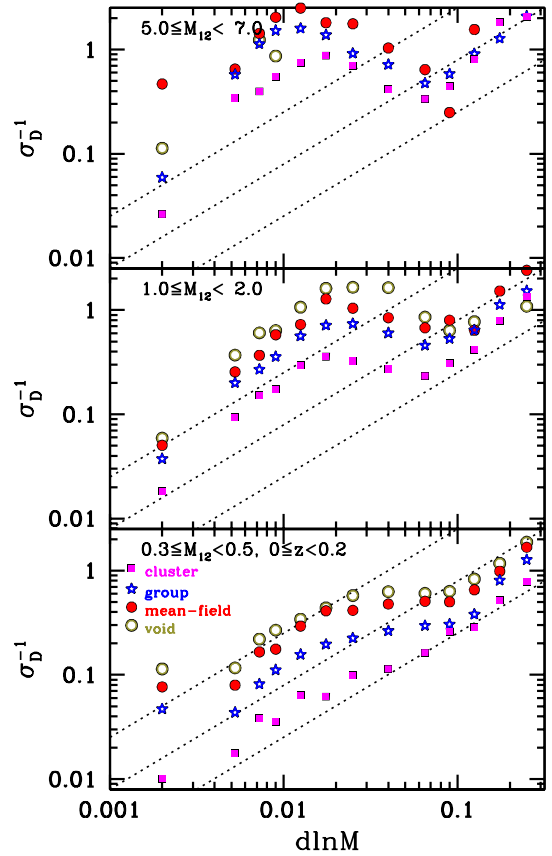}
\caption{
Changes in $\sigma_{\mathcal{D}}^{-1}$ and infall mass ratio,
$\Delta\log_{10} M$ (bottom), at $0\le z<0.2$ 
for different halo masses, $0.3\le M_{12}<0.5$ (bottom),
$1\le M_{12}<2$ (middle), and $5\le M_{12}<7$ ({\it top} panel).
The dotted lines show the linear relation between $\sigma_{{D}}^{-1}$ and $d\log_{10}M$.
}
\label{fig21}
\end{figure}

Figure \ref{fig21} shows the change in $\sigma_{{D}}^{-1}$ over a range 
of infall mass ratios at $z=0$.
The measured relation substantially deviates from a single power-law scaling but 
we can divide the range of $\Delta \log_{10}M$ into three regions:
$\Delta\log_{10}M < 0.03$ (quiet accretion), 
$0.03 < \Delta\log_{10}M \leq 0.1$ (strong accretion), and
$\Delta\log_{10}M > 0.1$ (major merging). 
Each region, except for the strong accretion mode, has its own power-law 
relation of $\sigma_D^{-1} \propto (\Delta\log_{10}M)^{1/2}$.
The major-merging mode and quiet accretion mode have different variances 
in that the major-merging mode has a higher value of $\sigma_c$.
This implies that the major merger mode has a relatively larger stochastic 
walk size than the quiet accretion (see Eq. \ref{gbm}).
The transition (or strong accretion) mode ($0.03\le \Delta\log_{10}M <0.1$) 
does not seem to satisfy the assumptions of the GBM model.
Now we conclude that 
even though there exist limitations in applying the GBM model 
to spin evolution, 
the spin walk is well described by the GBM model and, consequently,
the resulting spin distribution follows the log-normal function.

\section{Conclusions}\label{sec:concl}
We concluded that the spin distribution is well described by a random walk 
of angular momentum. 
Halos in lower density regions in recent epochs present some failures.
This may be due to correlated major-merging events 
(see further discussions in Appendix~\ref{appen1} and \ref{appen2}).
The log-normality of the spin distribution is found to be a consequence 
of the stochastic random walk, and most density samples 
show log-normal distributions of the halo spin. Only a small
departure is found in the cluster samples at $z=0$, mainly because
a fraction of halos are not yet relaxed.
We found that the angular momentum of a halo is likely to increase
after merging, and the spin distribution depends on the sample mass. 
If the sample mass is higher, then the average spin value decreases. 
The simulated log-normal distribution of the spin is well
recovered by the stochastic model with bimodal Gaussian distributions of 
angular momentum change.

In the standard $\Lambda$CDM model, the universe has undergone a recent 
accelerating expansion,
making it difficult to pull nearby material into a local shallow 
gravitational center.
Compared with the flat $\Omega_m=1$ model,
the $\Lambda$CDM model has prominent filamentary structures
lacking many small and faint structures that could be easily destroyed by the recent 
cosmic acceleration. 
This effect is even stronger
in void regions where the mass evolution was already halted at higher redshifts.
Therefore, the cosmological effects may leave clear evidence
in the spin evolution of void halos and mean-field halos at lower redshifts
($0.7\le \Delta\rho_{10}<$2; see the top-left panel of Fig. \ref{fig16}).

Because of the bimodal Gaussian shape of the angular momentum change in the 
stochastic model, 
one may raise the question of the possible existence
of another hidden parameter for subsampling merging events to measure $P(D)$.
By introducing another parameter,
a simpler distribution shape may be obtained.
A bimodal Gaussian shape could be obtained through a possible correlation among 
spin walks, which seems to be significant in void regions and in mean fields. 
This may explain the deviations in halo spin 
distributions from the log-normal distribution.

In this study, we directly adopt the simulated mass evolution for the halo mass 
change ($\Delta\log_{10}M$).
There are several models for the mass merging based on the EPS 
formalism (\cite{press74} for the classical 
and \cite{bower91, bond91, lacey93, kauffmann93,mo96, sheth99, 
sheth01,sheth02,vandenbosch02,hiotelis06,
zhang08, moreno08,neistein08,parkinson08,jiang14b}
for variant EPS models),
which can also provide us the probability of merging phases (major merger
or accretion) between time steps.
Since our stochastic model may discriminate between
the effects of accretion and major merging in spin distributions, we may
check whether the variant EPS models may produce proper probability 
distributions
of $\Delta M/M$ compared to simulated ones for a given time step.
In Appendix~\ref{epsmodel}, 
we show the roles of major merging and accretion in shaping the spin 
distribution in detail by adopting a toy model with fixed
the mass-merging ratio and how much the mass-merging history 
generated by the classical EPS model predicts different spin 
distributions from the $N$-body simulated ones.
 
\acknowledgments{
The authors thank Professor Changbom Park for many valuable comments on the 
dependence of spin evolution on the local environments.
J.K. appreciates the support of the Center for Advanced Computation
which provided the computing resources to run the simulations 
and analyze the massive halo data.
Y.-Y.C. was supported by a grant from Kyung Hee University in 2011 (KHU-20100179)
and the National Research Foundation
of Korea to the Center for Galaxy Evolution Research (No. 2010-0027910).
J.-E.L. was supported by the 2013 Sabbatical Leave Program of Kyung
Hee University (KHU-20131724).
S.S.K. was supported by the National Research Foundation grant funded by the Ministry of Science, ICT and Future Planning of Korea (NRF-2014R1A2A1A11052367).
}

\appendix

\section{Effect of $\alpha(z)$ on Spin Distribution}\label{sec:alpha}

$\alpha(z)$ deserves more investigation because its effect on 
spin distribution is not negligible. 
The cosmological contributions to log-spin change with a mass
increase in $\Delta\log_{10}M$ are 
$-0.008\lesssim \alpha(z)\Delta\log_{10}M <0$ in most redshift intervals
of merging data. So, even though its contribution is neglected, 
spin error is less than 1\%. 
But, the error accumulates over the number of merging steps 
and may substantially change the spin distribution.

\begin{figure}[tp]
\plotone{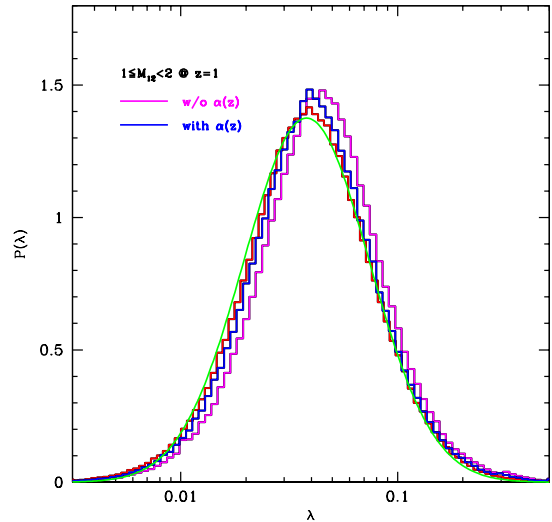}
\caption{
Effect of $\alpha(z)$ on the spin distribution at $z=1$.
The histograms colored in blue and magenta  are obtained with and without 
$\alpha(z)$ in Eq. \ref{cosmos}, respectively.
As references, the red histogram and a green curve are
the $N$-body spin distribution and its fitting function, respectively.
}
\label{fig_A1}
\end{figure}

Figure \ref{fig_A1} shows how the spin distribution changes
when the $\alpha(z)$ term is neglected in the random walk process. 
The generated spin distribution is substantially shifted
to higher $\lambda$, which demonstrates the sensitivity of
spin walk to such a very small value or possible error in fitting $P(D)$
and how large the sample size should be to suppress this kind of a noise effect. 
However, this does not guarantee that we may determine the cosmological model
from the measured spin distribution because 
the measured $P(D)$ in this analysis may differ in different cosmological models.

\section{Angle Alignment between Rotational and Orbital Angular Momenta}
\label{appen1}
\begin{figure}[tp]
\begin{center}
\subfigure[accretion]{ 
\includegraphics[width=232pt]{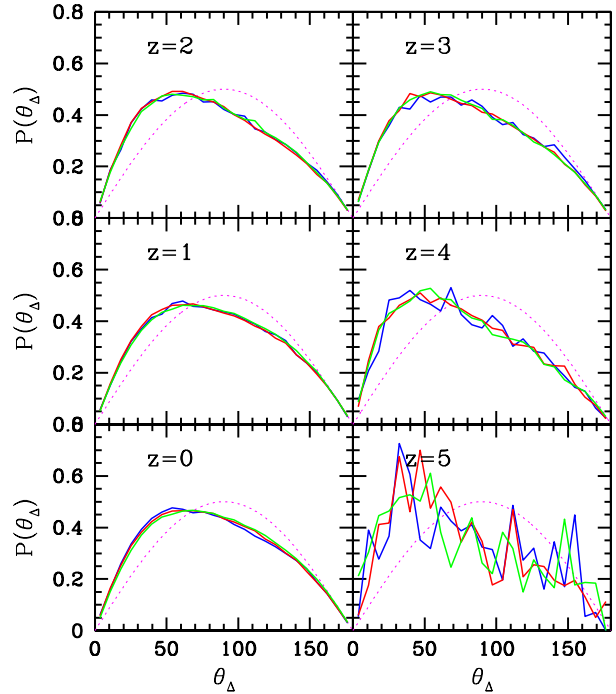}
}
\subfigure[merging]{
\includegraphics[width=232pt]{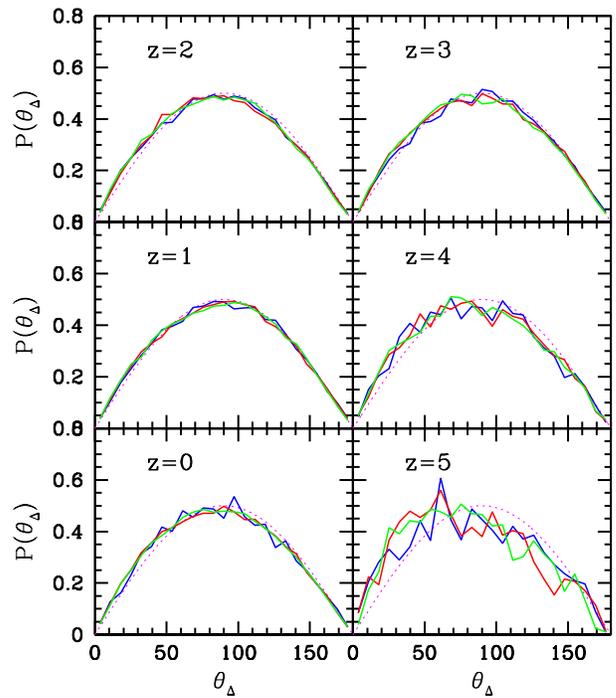}
}
\caption{ \label{fig_A2}
Angle alignment between $\boldsymbol{J}$ and $\boldsymbol{L}$ of 
accretion (a) in major-merging events and (b)
in halo samples of $0.03\le \lambda < 0.05$ and $1\le M_{12}<2$.
Counter-clockwise from the lower left panel, sample redshifts are $z=0$, 1, 2, 3, 4, and 5.
Each color-coded curve is measured from
the local-density sample of void (blue), mean-field (red), or
group (green) regions.
The magenta sine curve (dotted line) 
marks the random orientation.
}
\end{center}
\end{figure}

\begin{figure}[tp]
\begin{center}
\subfigure[accretion]{2
\includegraphics[width=232pt]{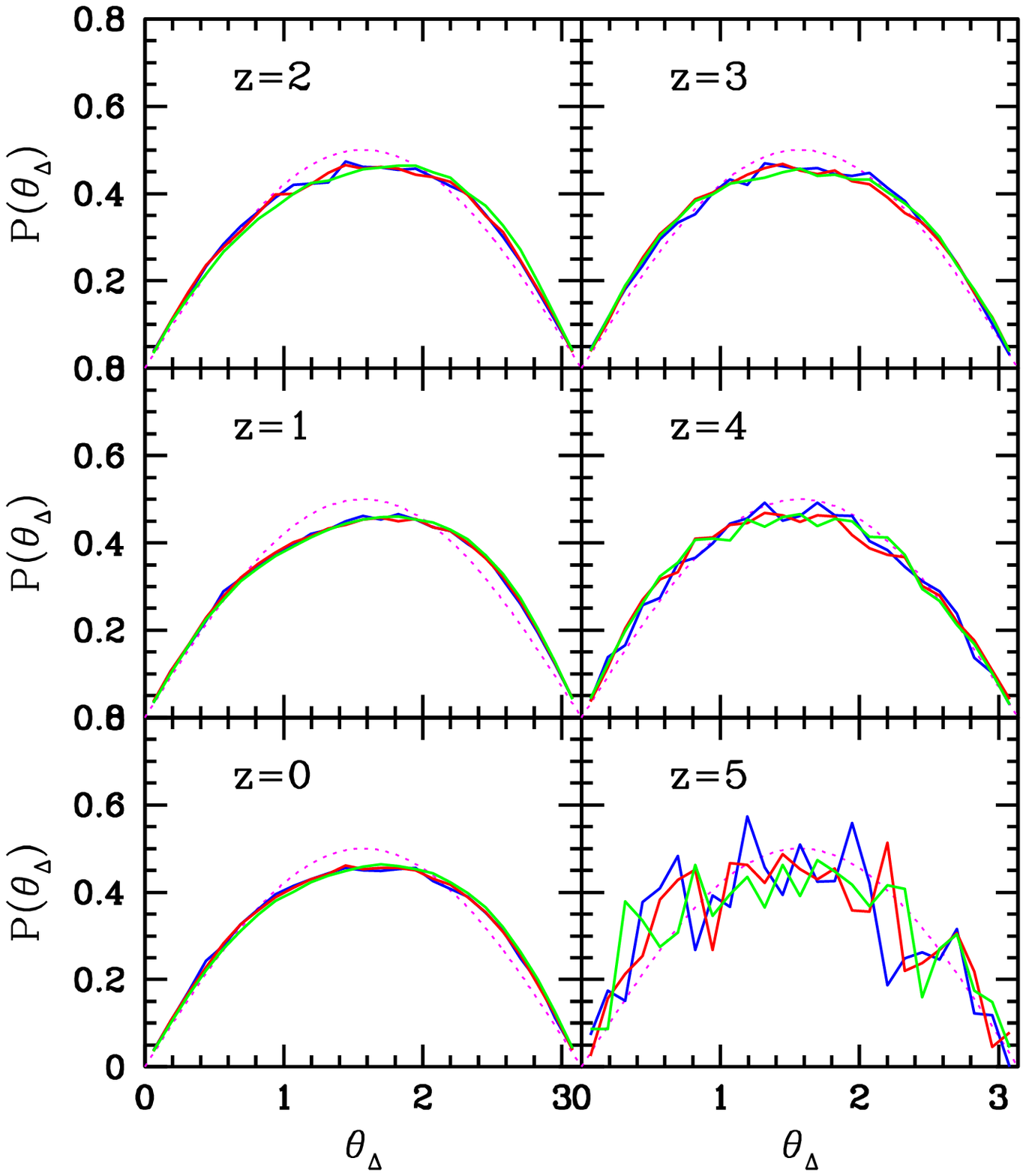}
}
\subfigure[merging]{
\includegraphics[width=232pt]{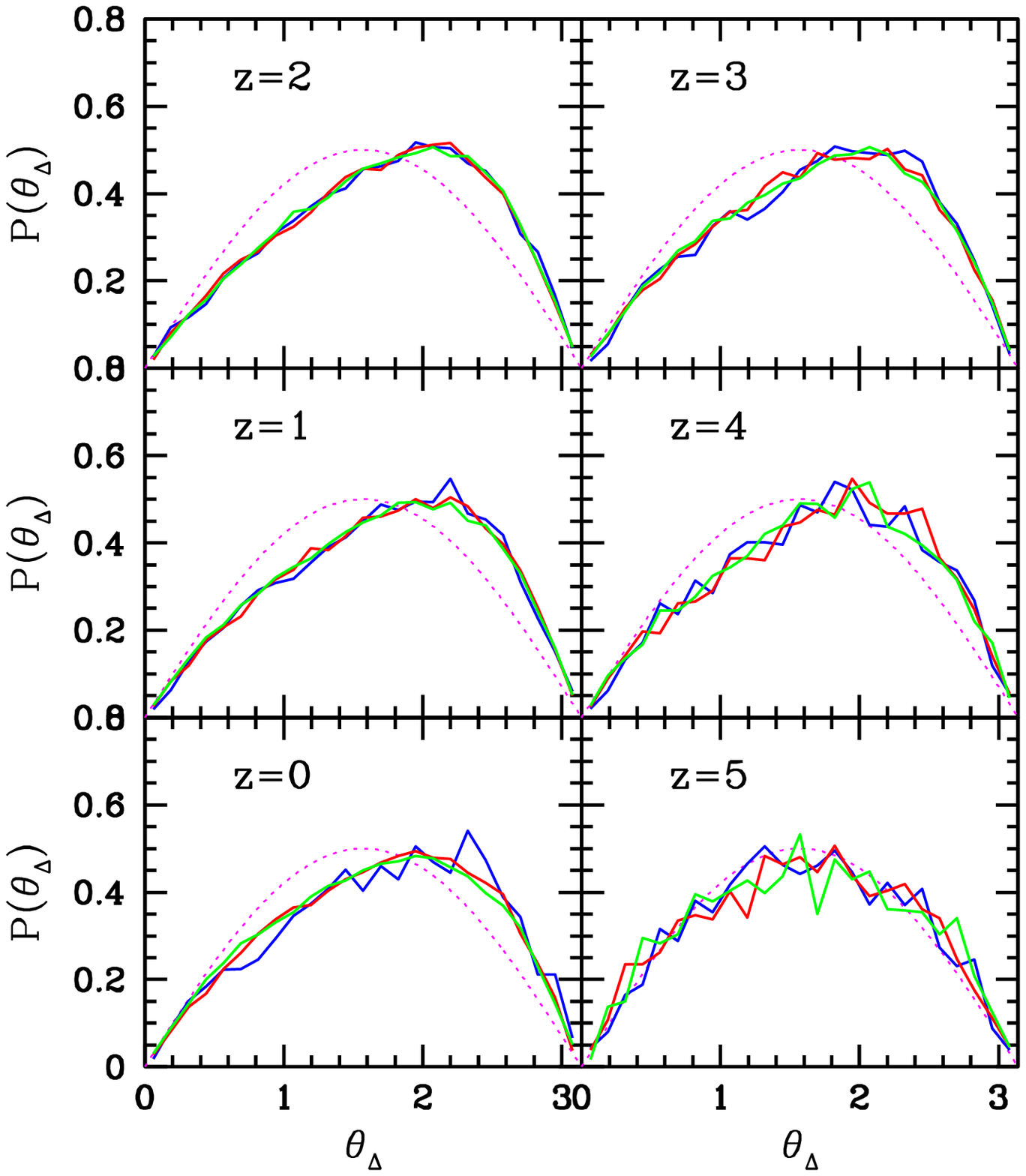}
}
\caption{ \label{fig_A3}
Same as Fig. \ref{fig_A2} but for slowly rotating samples of $0.01\le\lambda<0.03$.
}
\end{center}
\end{figure}

\begin{figure}[tp]
\begin{center}
\subfigure[accretion]{ 
\includegraphics[width=232pt]{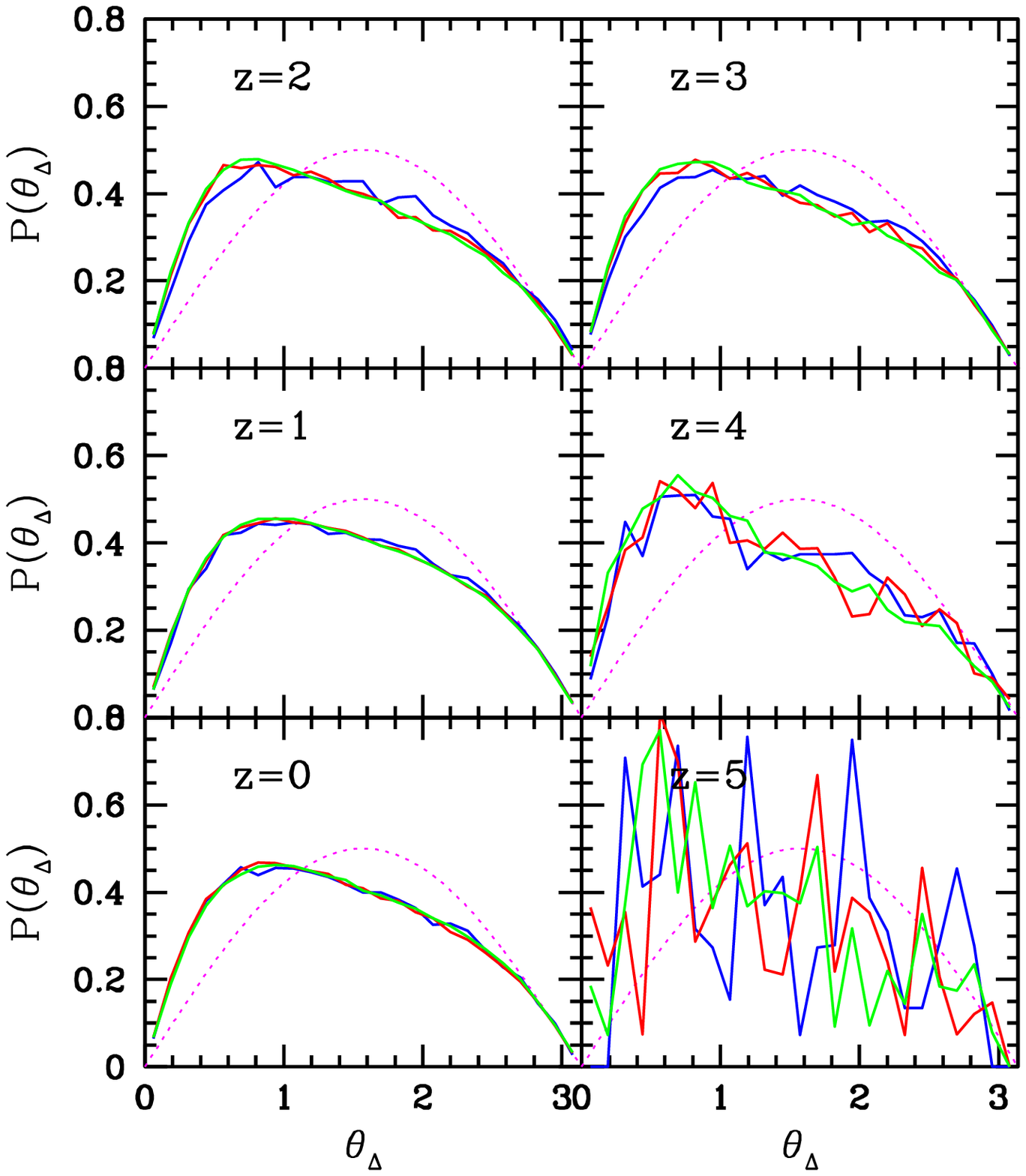}
}
\subfigure[merging]{
\includegraphics[width=232pt]{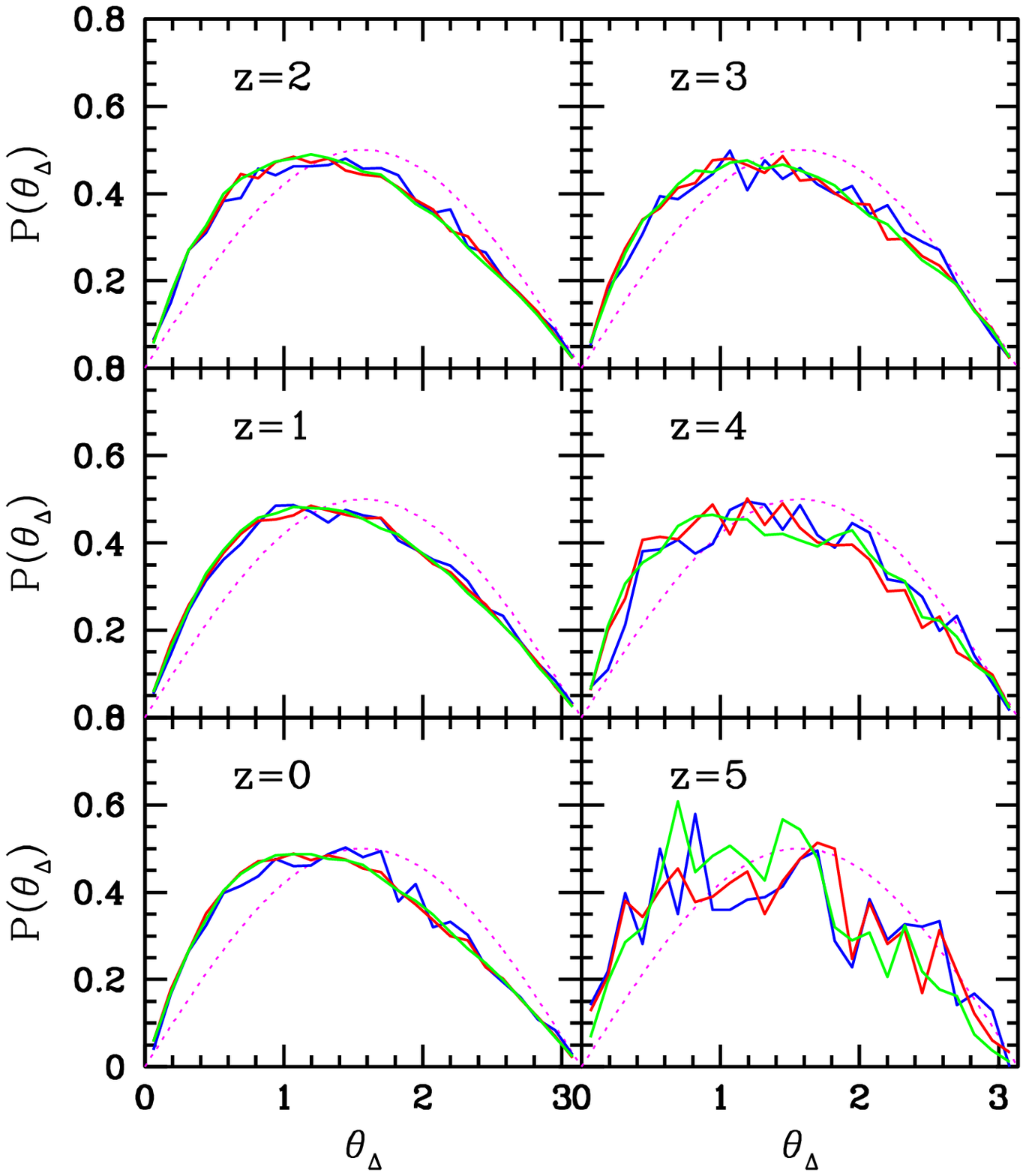}
}
\caption{ \label{fig_A4}
Same as Fig. \ref{fig_A2} but for fast-rotating samples of $0.05\le\lambda<0.3$.
}
\end{center}
\end{figure}

It would be interesting to examine the angle alignment
between the rotation ($\boldsymbol{J}$) of a halo and the orbital angular momentum 
(${\boldsymbol{L}}$) of infalling matter. 
If they are positively or negatively aligned, then
the rotational angular momentum 
of a halo is expected to increase or decrease, correspondingly. 
Therefore, this analysis is complementary to the work on the change in angular 
momentum ($D$).
In this section,
we study two infall modes: accretion ($0\le \Delta\log_{10}M <0.1$)
and major merging ($\Delta\log_{10}M\ge 0.1$).
The angle between two vectors is measured by
\begin{equation}
\theta_{\Delta} \equiv \cos^{-1} \left( { {\boldsymbol{J}} \cdot {\boldsymbol{L}} \over |{\boldsymbol{J}}||{\boldsymbol{L}}| } \right).
\end{equation}
Figures \ref{fig_A2}, \ref{fig_A3}, and \ref{fig_A4}
show the probability distribution of $\theta_{\Delta}$ in halo samples 
of $0.03\le\lambda<0.05$ (low-spi), 
$0.01\le\lambda<0.03$ (average-spin), and $0.05\le\lambda<0.3$ (
high-spin), respectively, for two mass-merging modes
(the accretion in the {\it left} and major merging in the {\it right panels}). 
In the accretion mode, the low-spin halos
(Fig. \ref{fig_A3}) are likely to have slightly
negative correlations with $\theta_{\Delta}$ (anti-aligned), 
while average-spin halos tend to have a slightly positive alignment,
and high-spin halos (Figs. \ref{fig_A2} and \ref{fig_A4})
have a strong positive alignment.
On the other hand, in major-merging mode,
the low-spin, average-spin, and high-spin samples show
negative, random, and positive correlations, respectively.

If two angular momentum vectors are anti-aligned, then a halo's rotational
angular momentum ($\boldsymbol{J}$) is reduced and, consequently, 
its spin value is lowered. 
Therefore, a halo with an average spin value is likely to have a slightly 
positive correlation to offset the negative drag term ($-5/3$),
while a high-spin or low-spin halo tends to have positively or 
negatively aligned mass infalls, respectively. 
Then, the alignment of two vectors and the value of a halo spin
are closely correlated with each other, producing a high spin value,
and vice versa.

\section{Angle Correlations between Mass Events}\label{appen2}

\begin{figure}[tp]
\epsscale{0.8}
\plotone{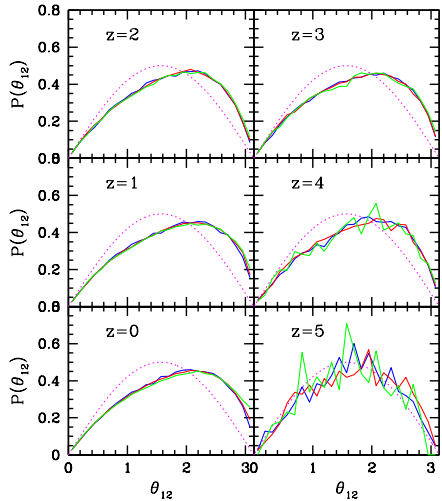}
\caption{ \label{fig_A5}
Angle correlations ($\theta_{12}$) of orbital angular momenta for
subsequent mergings in slow-rotating samples with $0.01\le\lambda<0.03$.
Each color-coded curve is measured from
the local-density sample of void (blue), mean-field (red), or
group (green) halos.
The magenta sine curve (dashed line) 
marks the average expectation of random orientations.
}
\end{figure}

\begin{figure}
\plotone{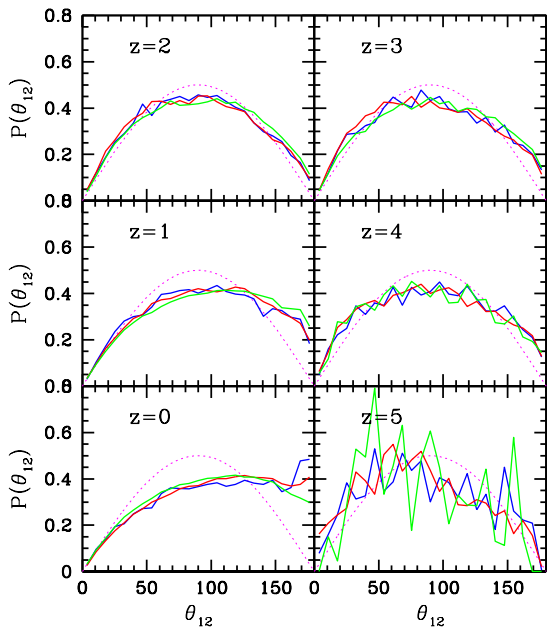}
\caption{ \label{fig_A6}
Same as Fig.~\ref{fig_A5} but for the mildly rotating sample with $0.03\le\lambda<0.05$.}
\end{figure}

\begin{figure}
\plotone{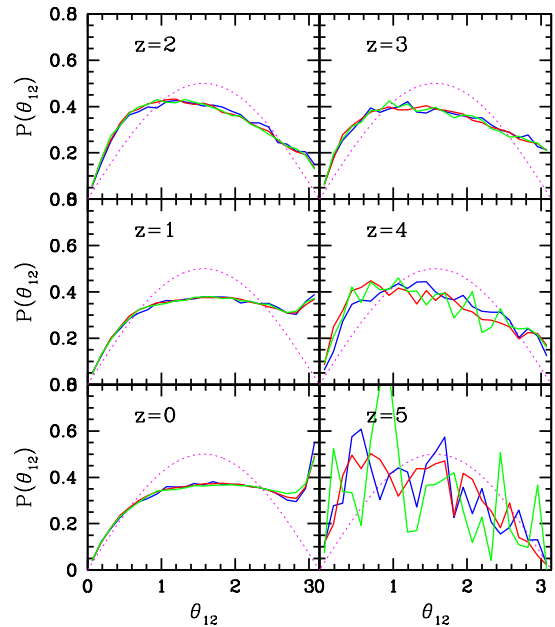}
\caption{ \label{fig_A7}
Same as Fig.~\ref{fig_A5} but for the fastly rotating sample with $0.05\le\lambda<0.3$.}
\end{figure}

This section discusses testing to determine 
whether or not the directions of the orbital angular momenta 
of infalling matter in two events are correlated.
The angle between the orbital angular momenta in the $i$th and $j$th events is 
defined as
\begin{equation}
\theta_{ij} \equiv \cos^{-1} \left( { {\boldsymbol{L}}_i \cdot {\boldsymbol{L}}_j \over |{\boldsymbol{L}}_i||{\boldsymbol{L}}_j| } \right),
\end{equation}
and we apply a 
weighting of $w_i=\Delta\log_{10} M$ placing more weight on major-merging pairs.
We only consider $j=i+1$ cases.
Figures \ref{fig_A5}--
\ref{fig_A7} show the directional correlations of the slowly, mildly, and 
fastly rotating halos, respectively.
In the figures,
there is no explicit dependency on environment.
Slow-rotating halos show a negative correlation (anti-aligned)
with regard to the direction of $\boldsymbol L$ between
two successive events, which explains why they are slowly rotating.
However, some interesting features can be found in the mildly and fastly rotating 
samples, which develop a strong anti-correlation at lower redshift ($z\le 1$;
bottom-left panel of Figs. \ref{fig_A6} and \ref{fig_A7}). 
This may curb the development of
spin as a whole and may explain the reason why void halos seem to 
have a non-Markovian walk of $D$.
Halos in denser regions are expected to experience frequent major mergers. 
Thus, even though there is anti-alignment between two steps, 
the many other major merger events violently erase the memory of 
previous anti-correlations.
Therefore, the superpositions of multiple anti-correlations are completely mixed
and may result in a random-like process. 
However, void halos tend to experience relatively few major mergers, and 
the effect of anti-alignment in a rare pair of merging events
may survive for a longer time.
This may explain why void halos are not properly described by the stochastic 
random walk of $D$.
We discuss the number of major mergers a halo may experience 
during its life time in Appendix \ref{append4}.

\section{Autocorrelations between Major Mergers}
\begin{figure}[tp]
\begin{center}
\subfigure[$1\le M_{12}<2$]{ 
\includegraphics[width=232pt]{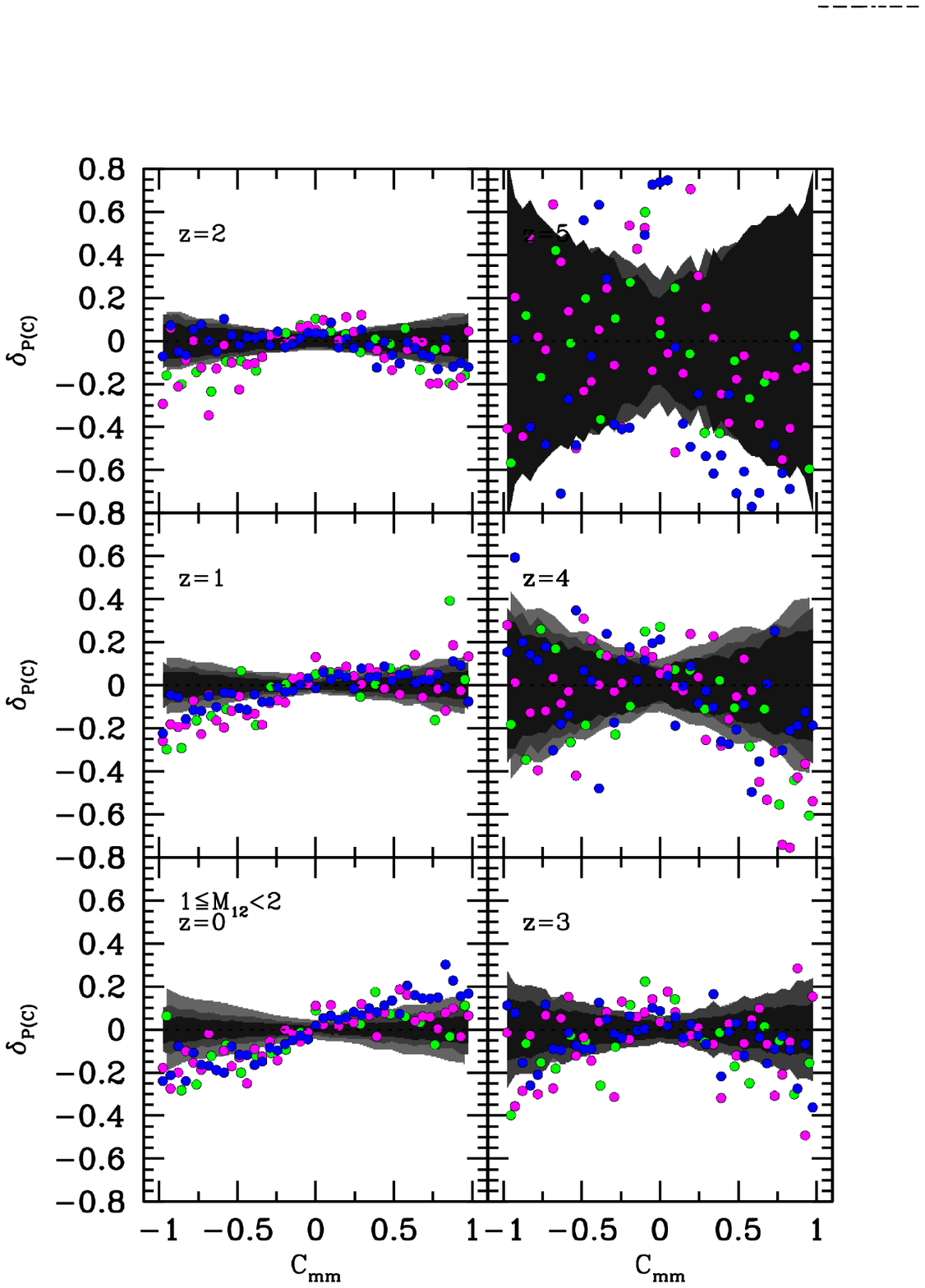}
}
\subfigure[$6\le M_{12}<10$]{
\includegraphics[width=232pt]{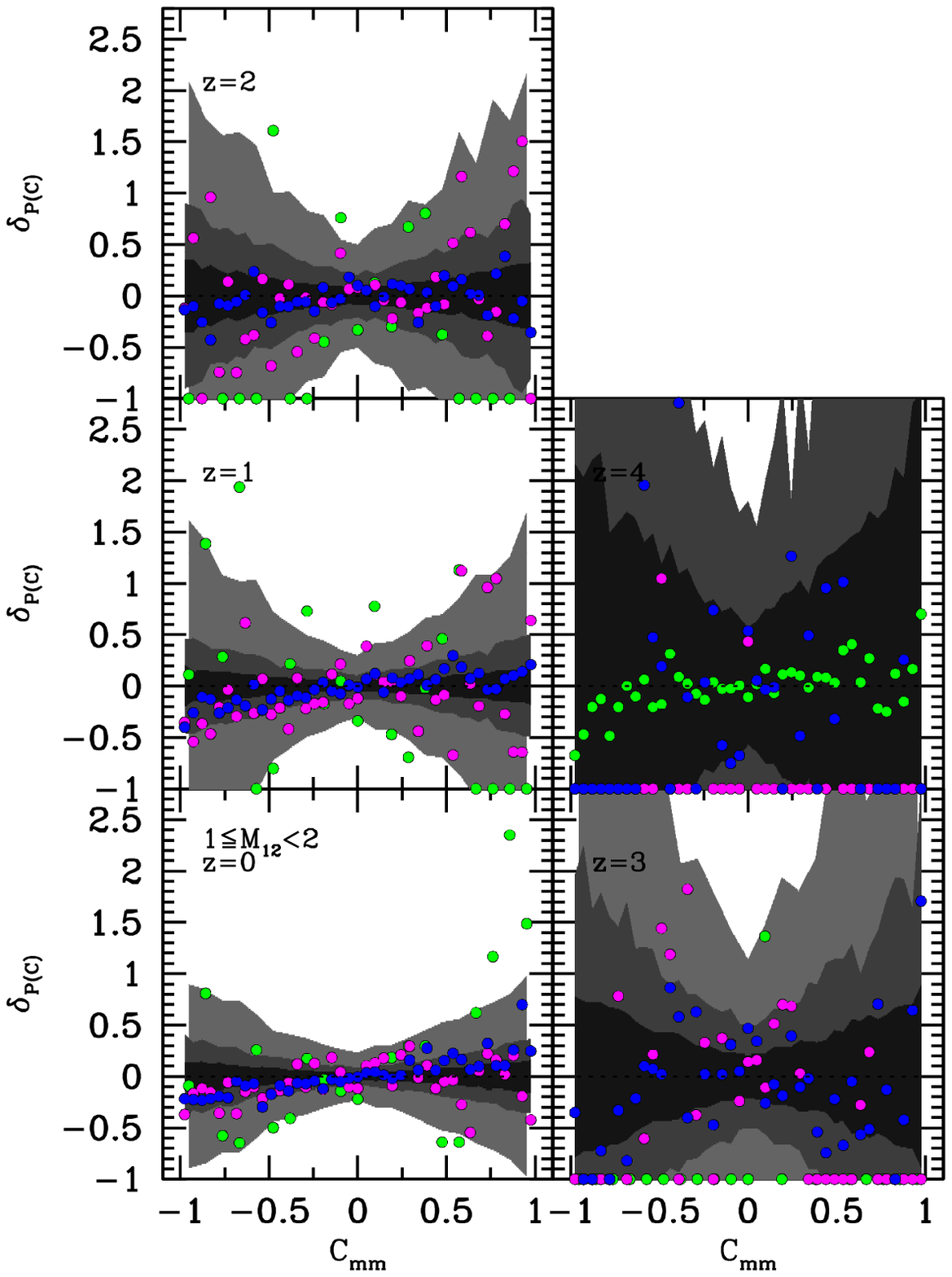}
}
\caption{ \label{fig_A8}
Autocorrelations between the major mergers at redshifts $z=0$, 1, 2, 3, 4, and 5
in the samples of $1\le M_{12}<2$ (a) and $6\le M_{12}<10$ (b).
}
\end{center}
\end{figure}

We estimate the autocorrelation between a pair of nearest major merger 
events to determine whether substantial amounts of 
correlation could be detected. 
In this case, we only consider autocorrelations between major merger events.
The deviations of autocorrelations between major mergers ($C_{\rm mm}$) 
are shown in Figure \ref{fig_A8} for several redshifts. 
In this figure, major merger autocorrelations 
at high redshifts ($z\ge2$) are equivalent to the
random distributions, while weak positive correlations are 
detected ($\simeq 1\sigma$) at lower redshifts.
More massive samples do not show any positive or negative correlations,
as seen in panel (b) of Figure \ref{fig_A8}.
Therefore, in this plot, 
we are sure that there is no substantial autocorrelation between major mergers.

\section{Number Distributions of Major Mergers}
\label{append4}
\begin{figure}[tp]
\begin{center}
\subfigure[$1\le M_{12}<2$]{ 
\includegraphics[width=232pt]{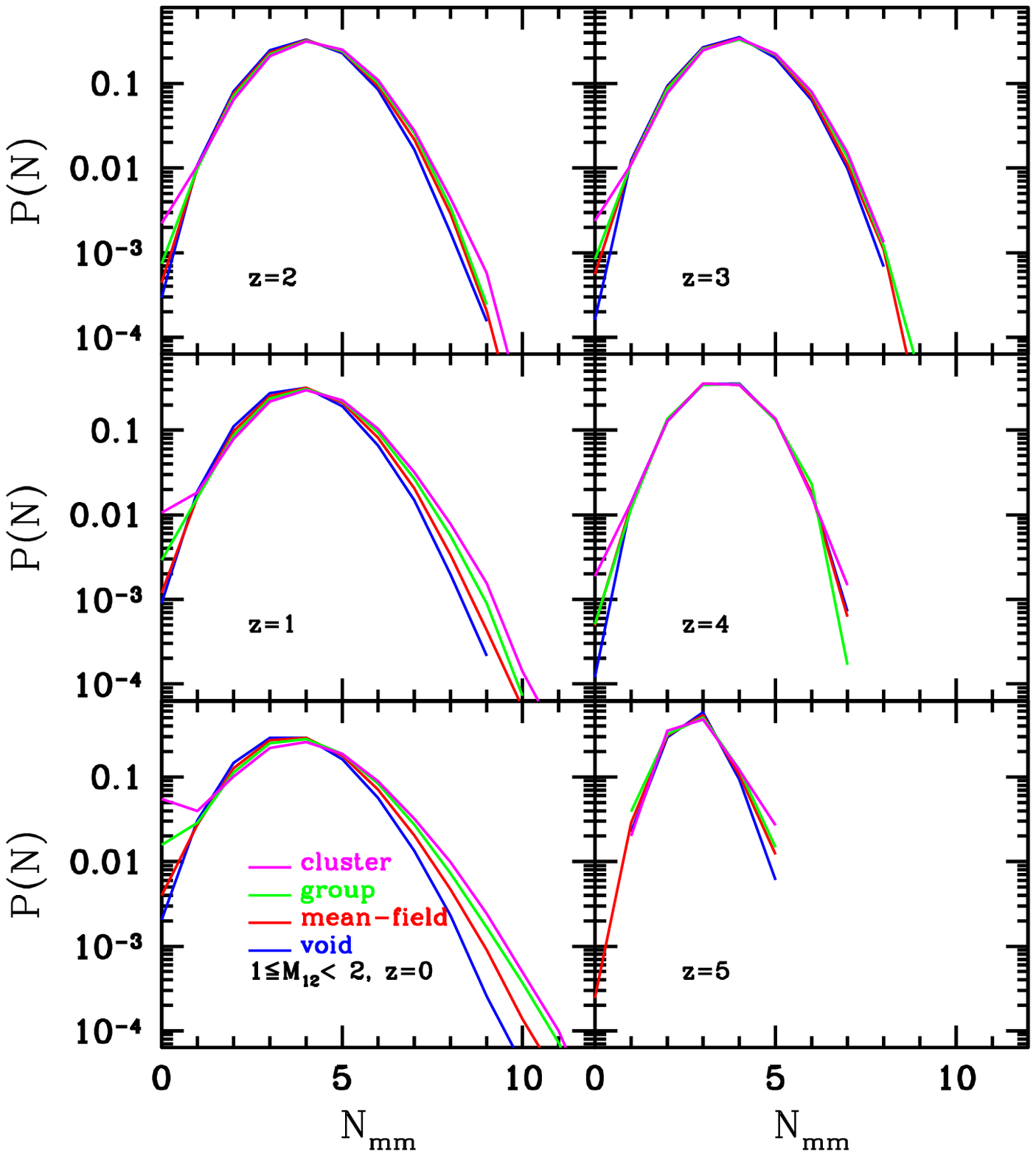}
}
\subfigure[$6\le M_{12}<10$]{
\includegraphics[width=232pt]{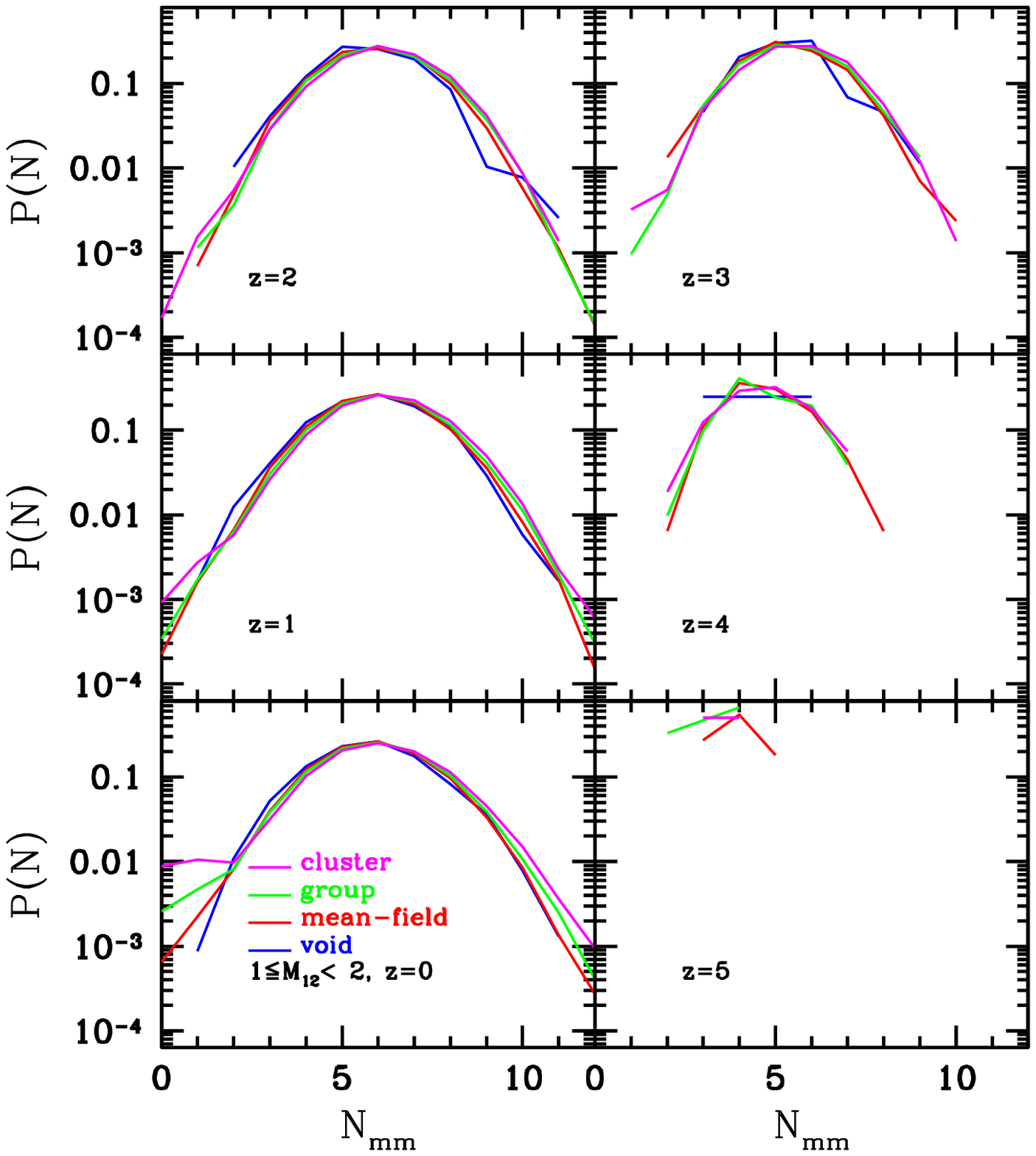}
}
\caption{\label{fig_A9}
Number distributions of major mergers in the history of a halo until $z=0$, 1, 2, 3, 4, and 5
measured from the samples of $1\le M_{12}<2$ (a) and $6\le M_{12}<10$ (b).
}
\end{center}
\end{figure}

We determine the distribution of major merger frequency in each halo 
sample at $z=0$ to confirm our argument that void halos experience fewer
major mergers, which may explain the deviations
of the stochastic model prediction from $N$-body spin distributions.
In Figure \ref{fig_A9}, we show the number of major mergers ($N_{\rm mm}$) 
per halo with masses $1\le M_{12}<2$ (a) or $6\le M_{12}< 10$ (b).
In the sample of $1\le M_{12}<2$, 
the number difference between halos in different environments is not
severe at higher redshift ($z\ge 2$),
while, at lower redshifts, less dense samples clearly show fewer major mergers. 
However, more massive samples (b) show nearly negligible deviations 
from each other, demonstrating noticeable differences only at $z=0$. 
The peak number of the $N_{\rm mm}$ distribution is four, while it is three at higher 
redshifts ($z\ge 4$).
Therefore, we conclude that the correlated random walk seen in void regions 
may be a joint effect of the major-to-major merger
autocorrelation and the smaller number of major mergers in less dense regions.
It is interesting to find that a significant fraction of halos in 
cluster regions do not experience a major merger event
during their lifetime (5\% of halos with $1\le M_{12} < 2$ and 1 \% 
of halos with $6\le M_{12}<10$).
Also, the fraction of cluster halos needed to prohibit a major merger 
event after $z=1$ is significantly higher than that of 
halos in lower density regions.

\section{Statistical Properties in the Spin Distribution}\label{appen5}
In this section, we present the means and standard deviations of 
spin distributions from $N$-body simulations and random generations 
together with fitting results to a log-normal function with
least $\chi^2$. 
See Tables~\ref{tab:fit1} and \ref{tab:fit2}.
The mean value of the halo spin increases with local density and redshift.
The difference between $N$-body and random simulations increases 
as local density decreases.

\begin{deluxetable*}{cc |ccc |ccc}
\tabletypesize{\footnotesize}
\tablecolumns{8} 
\tablewidth{0pt}
 \tablecaption{Spin Distributions for the Halo Mass Samples Used in
Figures~\ref{fig11}-\ref{fig13} \label{tab:fit1}}
 \tablehead{
 \multicolumn{2}{c|}{} & \multicolumn{3}{c|}{$N$-body Simulation}&  \multicolumn{3}{c}{Random Generation}\\
\cline{3-5} \cline{6-8}
Halo Mass & Redshift & 
$\langle \log_{10}\lambda \rangle$& $\sigma_{\log_{10}\lambda} $ 
&$\chi^2$/dof &
$\langle \log_{10}\lambda \rangle$& $\sigma_{\log_{10}\lambda} $ 
&$\chi^2$/dof
 }
 \startdata 
                & 0.0 & -1.386 & 0.336 & 0.00118 & -1.342 & 0.329& 0.00140 \\
$0.5\le M_{12} <0.8$ & 0.5 & -1.389 & 0.319 & 0.00076 & -1.374 & 0.308 & 0.00071 \\
                & 2.0 & -1.434 & 0.290 & 0.00034 & -1.429 & 0.289& 0.00037 \\
\hline
                & 0.0 & -1.407&  0.323& 0.00068& -1.358 &  0.326 & 0.00107 \\
$1\le M_{12} <2$ & 0.5 & -1.405 & 0.311 & 0.00037 & -1.381 &  0.309 & 0.00059 \\
                & 2.0 & -1.450 & 0.291 & 0.00029 & -1.427 & 0.294& 0.00031\\
\hline
             & 0.0 & -1.414 & 0.313 &0.00038& -1.366 & 0.324 & 0.00080 \\
$4\le M_{12} <6$ & 0.5 & -1.412 &  0.304 & 0.00028& -1.386 & 0.306 & 0.00049 \\
 & 2.0 & -1.474 & 0.291 & 0.00064 & -1.439 &  0.296 & 0.00062\\
\hline
& 0.0 & -1.415 & 0.309& 0.00027 & -1.367&  0.322& 0.00090 \\
$7\le M_{12} <10$ & 0.5 & -1.419 & 0.302& 0.00028 & -1.390 & 0.306& 0.00061 \\
 & 2.0 & -1.488&  0.292 &0.00086 &-1.446 & 0.304& 0.00110 \\
 \enddata
 \tablecomments{The mean and standard deviation of the halo spin from $N$-body simulation and random generation.
 Also, we list the least $\chi^2$ obtained during the log-normal fitting.}
\end{deluxetable*}

\begin{deluxetable*}{cc |ccc |ccc}
\tabletypesize{\footnotesize}
\tablecolumns{8}
\tablewidth{0pt}
 \tablecaption{Spin Distributions for the Halo Samples Used in
Figures~\ref{fig14}-\ref{fig17}\label{tab:fit2}}
 \tablehead{
 \multicolumn{2}{c|}{} & \multicolumn{3}{c|}{$N$-body Simulation}&  \multicolumn{3}{c}{Random Generation}\\
\cline{3-5} \cline{6-8}
Redshift&Environment &
$\langle \log_{10}\lambda \rangle$& $\sigma_{\log_{10}\lambda} $
&$\chi^2$/dof &
$\langle \log_{10}\lambda \rangle$& $\sigma_{\log_{10}\lambda} $
&$\chi^2$/dof
 }
 \startdata
\cutinhead{$1\le M_{12} < 2$}
     & void & -1.539 & 0.277 & 0.00108 & -1.440 & 0.297 & 0.00083 \\
$z=0$& mean-field & -1.486 & 0.284 & 0.00041 & -1.424 & 0.297 & 0.00057 \\
     & group & -1.410 & 0.311 & 0.00050 & -1.357 & 0.327 & 0.00083 \\
     & cluster & -1.287 & 0.375 & 0.00238 & -1.287 & 0.374 & 0.00130 \\
\hline
     & void & -1.544 & 0.273 & 0.00125 & -1.432 & 0.294 & 0.00059 \\
$z=1$& mean-field & -1.489 & 0.279 & 0.00053 & -1.435 & 0.291 & 0.00055 \\
     & group & -1.412 & 0.293 & 0.00028 & -1.391 & 0.299 & 0.00050 \\
     & cluster & -1.317 & 0.320 & 0.00058 & -1.332 & 0.320 & 0.00065 \\
\hline
     & void & -1.564 & 0.272 & 0.00168 & -1.457 & 0.290 & 0.00081 \\
$z=2$& mean-field & -1.522 & 0.277 & 0.00069 & -1.461 & 0.289 & 0.00043 \\
     & group & -1.454 & 0.285 & 0.00036 & -1.424 & 0.294 & 0.00038 \\
     & cluster & -1.369 & 0.298 & 0.00026 & -1.367 & 0.303 & 0.00043 \\
\cutinhead{$6\le M_{12} < 10$}
     & void & -1.574 & 0.268 & 0.01237 & -1.529 & 0.319 & 0.01626 \\
$z=0$& mean-field & -1.516 & 0.285 & 0.00112 & -1.459 & 0.318 & 0.00145 \\
     & group & -1.423 & 0.300 & 0.00036 & -1.378 & 0.315 & 0.00073 \\
     & cluster & -1.335 & 0.328 & 0.00072 & -1.290 & 0.347 & 0.00101 \\
\hline
     & void & -1.635 & 0.274 & 0.03620 & -1.585 & 0.305 & 0.02587 \\
$z=1$& mean-field & -1.549 & 0.280 & 0.00339 & -1.504 & 0.296 & 0.00257 \\
     & group & -1.447 & 0.288 & 0.00074 & -1.421 & 0.294 & 0.00067 \\
     & cluster & -1.360 & 0.298 & 0.00062 & -1.335 & 0.301 & 0.00090 \\\hline
     & void & -1.679 & 0.283 & 0.10107 & -1.589 & 0.349 & 0.10899 \\
$z=2$& mean-field & -1.613 & 0.270 & 0.01070 & -1.537 & 0.317 & 0.00637 \\
     & group & -1.520 & 0.281 & 0.00169 & -1.472 & 0.302 & 0.00155 \\     
     & cluster & -1.435 & 0.293 & 0.00154 & -1.384 & 0.306 & 0.00190 \\
 \enddata
 \tablecomments{Same as Table~\ref{tab:fit1}.}
\end{deluxetable*}

\section{Effect of MERGING MODEL on Spin Distribution}\label{epsmodel}

\begin{figure}
\plotone{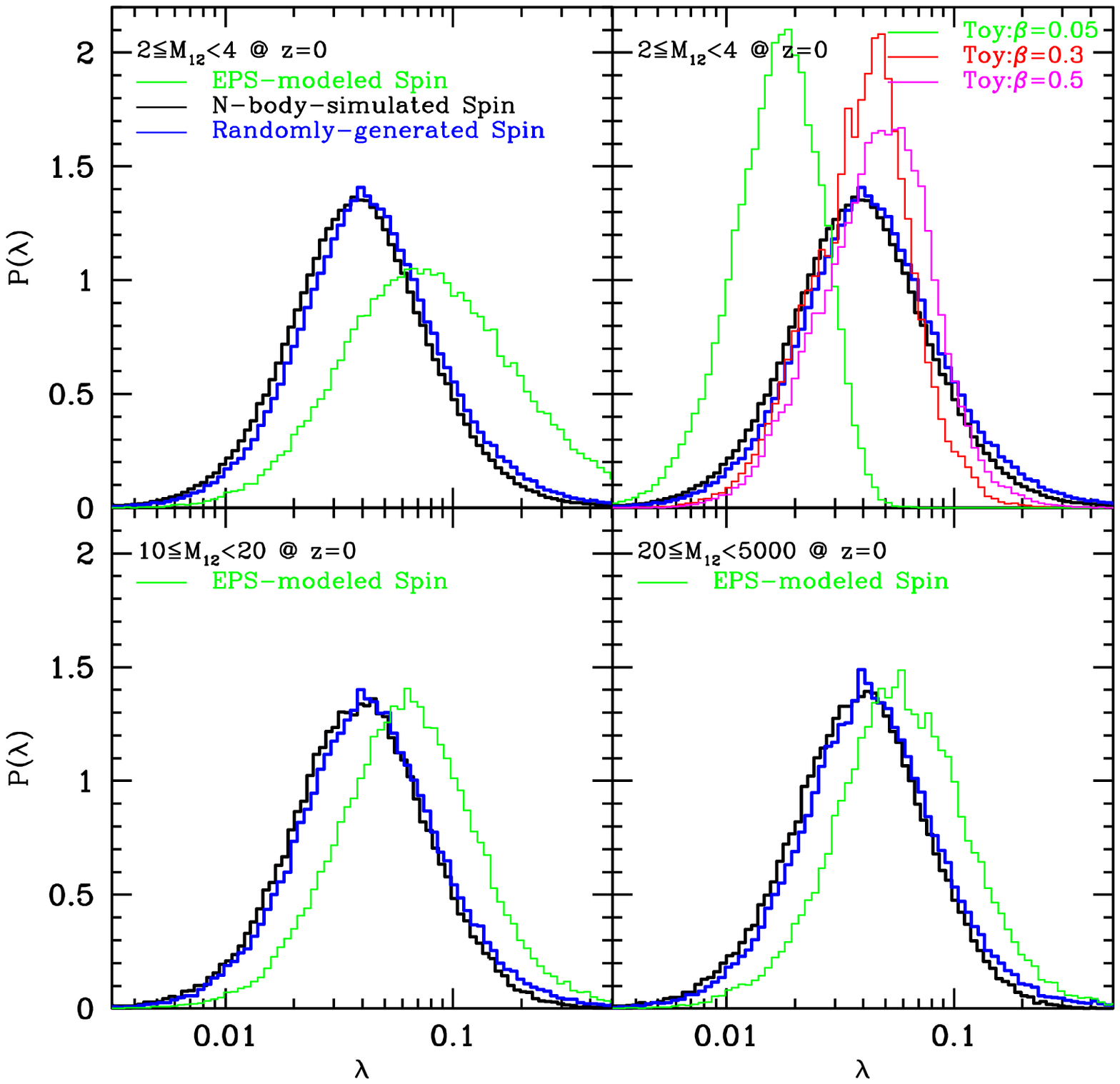}
\caption{ \label{fig_A10}
Spin distributions from the EPS model of the spherical top-hat collapse
\citep{lacey93};
top-left and bottom row panels)
and from the toy model with various $\beta$'s (top-right panel) at $z=0$.
}
\end{figure}
In this section, we demonstrate the effects of major merging and accretion 
on the spin distributions in a simple toy merging model.
The mass increase is parameterized as
\begin{equation}
\beta \equiv {\Delta M \over M},
\end{equation}
where $\beta$ is set constant over the halo mass evolution. 
The number of time steps is, therefore, determined
from the input value of $\beta$ for given initial and final halo masses.
In the top right panel of Figure \ref{fig_A10}, 
we observe that a higher value of $\beta$ makes the spin distribution move to
a higher $\lambda$. 
However, the actual mass growth is a spectrum of $\Delta M/M$ with 
a certain probability distribution and this shapes the
spin distributions found in simulations.

Next, we test whether the mass accretion histories derived from 
the classical extended Press-Schechter (EPS) formalism
may give a probability distribution
of $\Delta M/M$ comparable to the simulated ones.
We generate merging trees according to the EPS
merging model of the spherical top-hat collapse
\citep{lacey93}. 
The conditional probability of a progenitor to have a density 
contrast of $\delta_c(t_1)$ at the smoothing scale of $M_1$ is
\citep{jiang14b}
\begin{equation}
f_{\rm SC}(S_1,\omega_1 | S_0, \omega_0) = {1\over \sqrt{2\pi}}{\Delta \omega \over \Delta S^{3/2}}e^{-{\Delta\omega^2 \over 2 \Delta S}},
\end{equation}
where $\Delta S \equiv S_1-S_0$, $S_i\equiv \sigma^2(M_i)$,
$\Delta \omega \equiv \omega_1-\omega_0$, and $\omega_i \equiv \delta_c(t_i)$ for $t_0 > t_1$.
Using this equation, we generate the 100,000 random mass-merging trees and follow the stochastic spin changes
for several halo mass samples (see Fig. \ref{fig_A10}) at $z=0$.
In the figure, the EPS model predicts that the spin distributions will shift 
to higher values of $\lambda$, but
this tendency is smaller for larger mass samples.
This means that the spectrum of merger rates from the least accretion to most-massive major merger modeled by the classical EPS model
may be different from $N$-body results. 
Thus, the results suggest that it would be interesting to test 
whether other variant, and certainly more advanced, EPS models
that have been well known to provide halo mass functions consistent 
with simulations may also produce spin distributions similar
to those from $N$-body simulation by adopting this stochastic modeling.

\end{document}